\documentclass[lettersize,journal]{IEEEtran}
\usepackage{amsmath}

\usepackage{textcomp}
\usepackage{stfloats}
\usepackage{url}
\usepackage{verbatim}
\usepackage{graphicx}
\usepackage{cite}

\usepackage{graphicx}

\usepackage{enumitem}
\usepackage{pifont}
\usepackage{multirow}
\usepackage{amsmath}
\usepackage{cancel}
\usepackage{amssymb}
\usepackage{amsfonts}
\usepackage{array}
\usepackage{tcolorbox,cuted,lipsum}
\usepackage{tikz}
\usepackage[T1]{fontenc}
\usepackage{subcaption}
\usepackage{algorithmicx,float}
\usepackage[thinc]{esdiff}
\usepackage{xr}

\newtheorem{proposition}{\textbf{Proposition}}
\newenvironment{claimproof}[1]{\par\noindent{Proof:}\space#1}{\hfill $\blacksquare$}

\usepackage[linesnumbered,ruled]{algorithm2e}
\newenvironment{proc}[1][htb]
  {
   \begin{algorithm}%
  }{\end{algorithm}}

\usepackage{lineno,hyperref}
\modulolinenumbers[5]

\newtheorem{claim}{\textbf{Claim}}

\usepackage[linesnumbered,ruled]{algorithm2e}
\usepackage[export]{adjustbox}

\usepackage{hyperref}
\usepackage{float}

\begin{document}
%
\title{Strategic Analysis of Griefing Attack in Lightning Network}
%
%
%
%
\author{Subhra~Mazumdar,~Prabal Banerjee,~Abhinandan Sinha,\\ ~Sushmita~Ruj,~\IEEEmembership{Senior~Member,~IEEE}, and ~Bimal Kumar Roy 

\thanks{\textbf{Subhra Mazumdar} was with Cryptology and Security Research Unit, Indian Statistical Institute Kolkata, India. She is now with TU Wien and Christian Doppler Lab Blockchain
Technologies for the Internet of Things, Vienna, Austria. Email: subhra.mazumdar@tuwien.ac.at}
\thanks{\textbf{Prabal Banerjee} is with Cryptology and Security Research Unit, Indian Statistical Institute Kolkata, India and Polygon (previously Matic Network). Email: mail.prabal@gmail.com}
\thanks{ \textbf{Abhinandan Sinha} was with Economic Research Unit, Indian Statistical Institute Kolkata, India. He is now with Ahmedabad University, India. Email: abhinandan.sinha@ahduni.edu.in}
\thanks{\textbf{Sushmita Ruj} was with CSIRO’s Data61, Sydney, Australia. She is now with University of New South Wales, Sydney, Australia. Email: sushmita.ruj@unsw.edu.au} 
\thanks{\textbf{Bimal Kumar Roy} is with Applied Statistics Unit, Indian Statistical Insitute Kolkata, India. Email: bimal@isical.ac.in}}


%
%

\markboth{Journal of \LaTeX\ Class Files,~Vol.~14, No.~8, August~2021}%
{Shell \MakeLowercase{\textit{et al.}}: A Sample Article Using IEEEtran.cls for IEEE Journals}
%



\maketitle
\begin{abstract}
Hashed Timelock Contract (\emph{HTLC}) in Lightning Network is susceptible to a \emph{griefing attack}. An attacker can block several channels and stall payments by mounting this attack. A state-of-the-art countermeasure, Hashed Timelock Contract with Griefing-Penalty (\emph{HTLC-GP}) is found to work under the classical assumption of participants being either honest or malicious but fails for rational participants. To address the gap, we introduce a game-theoretic model for analyzing griefing attacks in \emph{HTLC}. We use this model to analyze griefing attacks in \emph{HTLC-GP} and conjecture that it is impossible to design an efficient protocol that will penalize a malicious participant with the current Bitcoin scripting system. We study the impact of the penalty on the cost of mounting the attack and observe that \emph{HTLC-GP} is \emph{weakly effective} in disincentivizing the attacker in certain conditions. To further increase the cost of attack, we introduce the concept of \emph{guaranteed minimum compensation}, denoted as $\zeta$, and modify \emph{HTLC-GP} into $\textrm{HTLC-GP}^{\zeta}$. By experimenting on several instances of Lightning Network, we observe that the total coins locked in the network drops to $28\%$ for $\textrm{HTLC-GP}^{\zeta}$, unlike in \emph{HTLC-GP} where total coins locked does not drop below $40\%$. These results justify that $\textrm{HTLC-GP}^{\zeta}$ is better than \emph{HTLC-GP} to counter griefing attacks.

\end{abstract}



\IEEEdisplaynontitleabstractindextext

%
\IEEEpeerreviewmaketitle

\section{Introduction}

Blockchain has redefined trust in the banking system. Transactions can be executed without relying on any central authority \cite{nakamoto2008bitcoin}. However, blockchain-based financial transactions cannot match traditional payment systems \cite{visa} in terms of throughput. Factors serving as the bottleneck are computation-overhead involved in verifying transactions, and expensive consensus mechanism \cite{croman2016scaling}. Layer 2 solutions \cite{gudgeon2019sok} have been developed on top of blockchain to address these shortcomings. One of the solutions, \emph{payment channels} \cite{decker2015fast} are widely deployed and quite simple to implement. Several interconnected payment channels form a payment channel network or PCN. The network is used for executing several transactions between parties not directly connected by a channel, without recording any of them in the blockchain.

  \begin{figure}[!ht]
    \centering
    \includegraphics[width=7cm]{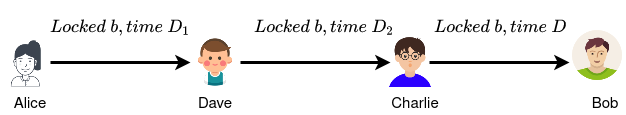}
    \caption{Formation of contract, forwarding conditional payment from \emph{Alice} to \emph{Bob}}
    \label{bob1}
\end{figure}


\textcolor{black}{Lightning Network is the most popular Bitcoin-based PCN \cite{poon2016bitcoin}. A payer can securely transfer funds to an intended recipient in the network by using a Hashed Timelock Contract or \emph{HTLC}. It is a form of conditional payment where a payment succeeds contingent on the knowledge of the preimage of a hash value. For example, in Fig. \ref{bob1}, \emph{Alice} intends to transfer $b$ coins to \emph{Bob} via channels \emph{Alice-Dave, Dave-Charlie}, and \emph{Charlie-Bob}. \emph{Bob} generates a hash $H=\mathcal{H}(x)$ and shares it with \emph{Alice}. The latter forwards a conditional payment to \emph{Dave}, locking $b$ coins for time $D_1$. \emph{Dave} forwards the payment to \emph{Charlie}, locking $b$ coins for $D_2$ units of time. Finally, \emph{Charlie} locks $b$ coins with \emph{Bob} for time $D$, where $D_1>D_2>D$. To claim $b$ coins from \emph{Charlie}, \emph{Bob} must release $x$ within the time $D$. If \emph{Bob} does not respond, then \emph{Charlie} withdraws the coins locked from the contract by going on-chain and closing the channel after the timeout period. However, \emph{Bob} manages to lock $b$ coins in each of the preceding payment channels for the next $D$ units \cite{egger2019atomic}. This is an instance of griefing attack \cite{robinson2019htlcs}. An empirical analysis \cite{bank, lu2020general} shows that a griefing attack reduces the liquidity of Lightning Network, and it has been used for eliminating specific edges in PCN \cite{rohrer2019discharged}.}



\begin{figure}[!ht]
    \centering
    \includegraphics[width=8cm]{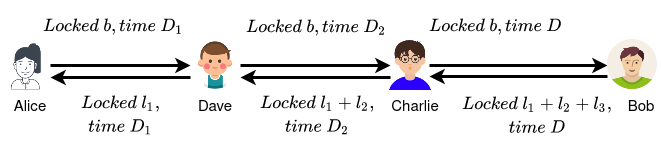}
    \caption{Off-chain contract formation in \emph{HTLC-GP}}
    \label{bob2}
\end{figure}

\textcolor{black}{A payment protocol Hashed Timelock Contract with Griefing Penalty or \emph{HTLC-GP} \cite{agrief} was proposed to counter griefing attack. The off-chain contract formation in \emph{HTLC-GP} is illustrated in Fig. \ref{bob2}. \emph{Dave} has to lock $l_1$ coins as a guarantee against $b$ coins locked by \emph{Alice}, for a period of $D_1$ units. If \emph{Dave} responds within $D_1$, he claims $b$ coins and unlocks $l_1$ coins. If he fails to respond, \emph{Alice} goes on-chain, closes the channel, unlocks $b$ coins, and gets the compensation from \emph{Dave}. Similarly, \emph{Charlie} has to lock $l_1+l_2$ coins for a period of $D_2$ units as a guarantee against $b$ coins locked each by \emph{Dave} and \emph{Alice}. \emph{Bob} locks $l_1+l_2+l_3$ coins for a period of $D$ units as a guarantee against $b$ coins locked by \emph{Charlie}, \emph{Dave} and \emph{Alice}. The drawback of \emph{HTLC-GP} is that it does not consider an attacker to be rational. If \emph{Bob} intends to mount griefing attack, he will choose a strategy that avoids paying any penalty. He will resolve the payment off-chain with \emph{Charlie} just before the contract's locktime elapses. In this way, he manages to lock \emph{Alice, Dave} and \emph{Charlie}'s coins without compensating them. We realize that the underlying punishment mechanism in \emph{HTLC-GP} must be argued from a game-theoretic point of view and not from the cryptographic aspect.}

\subsection{Contributions}
We make the following contributions in our paper:\\
\textcolor{black}{(i) This is the first attempt to propose a two-player game-theoretic model for analyzing griefing attacks in \emph{HTLC}. The first player receives a conditional payment and makes a decision of forwarding the same to the second player based on the belief that the latter may be \emph{corrupt} or \emph{uncorrupt.}}\\
(ii) We use the same game model to analyze the griefing attacks in \emph{HTLC-GP} \cite{agrief}. \textcolor{black}{From the analysis, we conjecture that it is impossible to design an effective countermeasure without changing the Bitcoin scripting system.}\\
(iii) \textcolor{black}{We analyze the impact of the penalty on the attacker's behavior and infer that \emph{HTLC-GP} is \emph{weakly effective} in countering the attack in certain conditions.}\\
(iv) \textcolor{black}{We introduce the concept of \emph{guaranteed minimum compensation}, $\zeta$, and propose a protocol, $\textrm{HTLC-GP}^{\zeta}$, for disincentivizing griefing attack.}\\
(v) \textcolor{black}{Our experimental analysis shows that the total coins locked by the attacker drops to $28\%$ when the guaranteed minimum
compensation is $2.5\%$ of the
transaction amount and the maximum allowed path length is set to $10$ in $\textrm{HTLC-GP}^{\zeta}$. This quantity is $27\%$ less
than the coins locked in \emph{HTLC-GP}, proving that $\textrm{HTLC-GP}^{\zeta}$ is far more effective than \emph{HTLC-GP} in countering the griefing attack. The code is provided in GitHub\footnote{\url{https://github.com/subhramazumdar/Strategic_Analysis_Griefing}}}.



\subsection{Organization}
Section \ref{background} discusses the background and Section \ref{related} discusses the state-of-the-art. In Section \ref{game1}, we propose a game-theoretical model of griefing attack in \emph{HTLC}. We discuss an existing countermeasure for griefing attack, \emph{HTLC-GP}, in Section \ref{countermeasure} and use the game model discussed in Section \ref{game1} to analyze the griefing attacks in \emph{HTLC-GP} in Section \ref{gpanalysis}. In Section \ref{guarantee}, we propose the concept of \emph{guaranteed minimum compensation}, denoted as $\zeta$. We modify \emph{HTLC-GP} into $\textrm{HTLC-GP}^{\zeta}$ by incorporating the concept of minimum compensation, and discuss in Section \ref{modify}. Experimental analysis of the effectiveness of \emph{HTLC-GP} and $\textrm{HTLC-GP}^{\zeta}$ is provided in Section \ref{experiment}. \textcolor{black}{Section \ref{scale} discusses how scalable is $\textrm{HTLC-GP}^{\zeta}$ compared to the state-of-the-art.} Finally, we conclude the paper in Section \ref{conclusion}.


\section{Background}
\label{background}

\subsubsection*{(i) Lightning Network}

It is a \emph{layer 2} solution for scalability issues in Bitcoin blockchain \cite{poon2016bitcoin}. It is modeled as a bidirected graph $G:=(V,E)$, where $V$ is the set of nodes and $E \subseteq V \times V$ is the set of payment channels opened between a pair of nodes. Every node charges a processing fee for processing payment requests, defined by a function $f$, where $f: \mathbb{R}^{+}\rightarrow \mathbb{R}^+$. Each payment channel $(U_i,U_j) \in E$ has an initial capacity $locked(U_i,U_j)$, denoting the amount deposited by $U_i$ in the channel. $t_{i,j}$ is the timestamp at which the channel was opened. In the context of the Bitcoin blockchain, this will be the block height. $T$ is the expiration time of the channel $id_{i,j}$, i.e., once a channel is opened, it remains active till time $T$ units\footnote{Each channel in Lightning Network has an infinite lifetime. However, we assume an upper bound on the channel lifetime for our analysis. Setting channel expiration time has been used in the literature as well \cite{malavolta2019anonymous}}. $remain(U_i,U_j)$ signifies the residual amount of coins $U_i$ can transfer to $U_j$ via off-chain transactions. $M$ denotes the average fee for mining a Bitcoin transaction.

\subsubsection*{(ii) Hashed Timelock Contract}
It is used for forwarding conditional payments across parties not directly connected by the payment channel \cite{poon2016bitcoin}. Suppose a payer $U_0$ wants to transfer funds to a payee $U_n$ via an $n$-hop route $P=\langle U_0,U_1,U_2,\ldots,U_n \rangle$ in the network. $U_n$ creates a condition $Y$ defined by $Y=\mathcal{H}(x)$ where $x$ is a random string and $\mathcal{H}$ is a cryptographic hash function \cite{rogaway2004cryptographic}. $U_n$ shares the condition $Y$ with $U_0$. The latter uses $Y$ for conditional payment across the whole payment path. Between any pair of adjacent nodes $(U_{i-1},U_{i})$ in $P$, where $i \in [1,n]$, the hashed timelock contract is defined by $HTLC(U_{i-1},U_{i},Y,b,t_{i-1})$, where $t_{i-1}=t_i+\Delta$, where $\Delta$ is the worst-case confirmation time for a transaction to get confirmed on-chain. The contract implies that $U_{i-1}$ locks $b$ coins in the off-chain contract. The coins locked can be claimed by the party $U_{i}$ only if it releases the correct preimage $x': Y=\mathcal{H}(x')$ within time $t_{i-1}$. If $\mathcal{H}$ is a collision-resistant hash function, then $x'\neq x$ with negligible probability. If $U_{i}$ doesn't release the preimage within $t_{i-1}$, then $U_{i-1}$ settles the dispute on-chain by broadcasting the transaction. The channel between $U_{i-1}$ and $U_{i}$ is closed and $U_{i-1}$ unlocks the coins from the contract. If both the parties mutually decide to settle off-chain then $U_{i}$ either releases the preimage and claims $b$ coins from $U_{i-1}$. If $U_i$ decides to reject the payment then $b$ coins are refunded to $U_{i-1}$.

\subsubsection*{(iii) Dynamic Games of Incomplete Information or Sequential Bayesian Games}
In this class of games, players move in sequence, with at least one player being uncertain about another player's payoff. We define a belief system and a player's behavioral strategy to approach these games. A \emph{type space} for a player is the set of all possible types of that player. A \emph{belief system} in a dynamic game describes the uncertainty of that player of the types of the other players. A \emph{behavioral strategy} of a player $i$ is a function that assigns to each of $i 's$ information set a probability distribution over the set of actions to the player $i$ at that information set, with the property that each probability distribution is independent of every other distribution. A dynamic game of incomplete information \cite{gibbons1992dynamic} is defined as a tuple that consists of (i) a set of players $\mathcal{I}$; (ii) a sequence of histories $H^m$ at the $m^{th}$ stage of the game, each history assigned to one of the players (or to Nature/Chance); (iii) an information partition. The partition determines which of histories assigned to a player are in the same information set; (iv) a set of pure strategies for each player $i$, denoted as $S_i$; (v) a set of types for each player $i: \theta_i  \in \Theta_i$; (vi) a payoff function for each player $i: u_i(s_1,s_2,\ldots,s_l,\theta_1,\theta_2,\ldots,\theta_l)$; (vii) a joint probability distribution $p(\theta_1,\theta_2,\ldots,\theta_l)$ over types.
\emph{Perfect Bayesian equilibrium} \cite{fudenberg1991perfect} is used to analyze dynamic games with incomplete information.


\section{Related works}
\label{related}
 
Few existing works analyze the payments executed in Lightning Network from a game-theoretic point of view but none of them have analyzed the griefing attack. \textcolor{black}{Zappala et al. \cite{zappala2020game}  proposed a framework for formally characterizing the robustness of blockchain systems in the presence of Byzantine participants}. The authors have defined \emph{HTLC} as a game between three participants. However, it is assumed that an \emph{HTLC} can either be accepted or rejected but there is no discussion on griefing. \textcolor{black}{Rain et al. \cite{rain2021towards} discusses the shortcoming of the game model of multi-hop payment proposed in \cite{zappala2020game}}. They have improved the model and analyzed \emph{wormhole attacks} in the network but the work does not discuss the griefing attack.

In \cite{xu2020game}, a game-theoretic analysis of atomic cross-chain swaps using \emph{HTLC} has been provided. They have studied the impact of token price volatility on the strategic behavior of the participants initiating the swap and suggested the use of collateral deposits to prevent parties from canceling the swap. \textcolor{black}{Han et al. \cite{han2019optionality} proposed use of premium for fairness in the atomic cross-chain swap. This is the first paper to introduce penalty as a countermeasure for countering the griefing attack in the context of atomic swap.} However, the paper lacks a detailed analysis of how the introduction of premiums might ensure a faster exchange of assets. Also, the premium calculated is not a function of the assets locked in the off-chain contract and its timeout period. This might lead to under-compensation of victims. Our work is the first to propose a game model for analyzing griefing attack in Lightning Network. Several countermeasures like upfront payments \cite{russell}, proof-of-Closure of channels \cite{zmn}, etc were proposed for mitigating the work. However, they are not effective because the malicious node does not lose anything in the process. Hashed Timelock Contract with Griefing Penalty or \emph{HTLC-GP} \cite{agrief} claims to address the above shortcomings and disincentivize griefing attacks. We use our proposed game model to study the effectiveness of \emph{HTLC-GP} and suggest appropriate modifications to the protocol.

\section{Analysis of griefing attack in HTLC}
\label{game1}


\subsection{System Model}
\label{sys-assumption1}
 \begin{table}[!ht]
\begin{center}
\scalebox{0.7}{
  \begin{tabular}{|c |c |} 
    \hline
 Notation &Description \\
 \hline
 
$G:=(V,E)$ &A bidirected graph representing the Lightning Network\\
$V$   &Set of nodes in Lightning Network\\
$E$   &Set of payment channels in Lightning Network, $E \subset V \times V$\\

  $\alpha$ & Amount to be transferred from sender $S$ to receiver $R$\\
$P$ &Path connecting $S$ to $R$\\
  $\kappa$ & Length of the path $P$ in $G$ connecting payer $S$ to payee $R$. \\
$n$ &Maximum allowed path length for payment, where $n\in \mathbb{N}, \kappa\leq n$\\
  
  $U_i \in V, i \in [0,\kappa]$ &Nodes in $P, (U_{i-1},U_{i}) \in E$, where $U_0=S$ and $U_{\kappa}=R$.\\
  
  $id_{i,j}$ & Identifier of channel $(U_i,U_j)$\\ 
  $T$ &Lifetime of a channel \\
$t_{i,j}$ & The time at which the channel between $U_i$ and $U_j$ was opened \\

  $locked(U_i,U_{j})$ &Amount of funds locked by $U_{i}$ in the payment channel $(U_i,U_{j})$ while channel opening.\\
  $remain(U_i,U_j)$ &Net balance of $U_i$ that can be transferred to $U_j$ via off-chain transaction \\
 $f(\alpha)$ &Processing fee charged by a node for forwarding $\alpha$ coins to its neighbor\\
 $\lambda$ &Security Parameter\\
 $\mathcal{H}\{0,1\}^*\rightarrow \{0,1\}^\lambda$ & Standard Cryptographic Hash function\\
 $\Delta$ &Worst-case confirmation time when a transaction is settled on-chain\\
  $t_i$ &HTLC Timeout period in channel $(U_i,U_{i+1}), i \in [0,\kappa-1]$\\
 $D$ &Least HTLC Timeout period (or $t_{\kappa-1}$)\\
 $L$ &Bribe offered per attack\\
 $I_{t,\alpha}$ &Compensation offered for keeping $\alpha$ coins unutilized for the next $t$ units of time\\
 $r_U$ &Rate of payments processed by node $U$ per unit time\\
$O(r_U,t,val)$ & Opportunity cost of a node $U$ for next $t$ units of time, also denoted as $o_{U}^{t,val}$\\
 
$t_{contract\_initiate}$ &Timestamp at which off-chain contract got initiated\\
 $M$ & Average fee charged for mining a transaction\\
 
 $\Gamma_{HTLC}$ &Extensive form of a 2-party sequential Bayesian game in \emph{HTLC}\\
$\gamma$ &Rate of griefing penalty (per minute) \\

 $\Gamma_{HTLC-GP}$ &Extensive form of a 2-party sequential Bayesian game in \emph{HTLC-GP}\\
$\zeta$ & Guaranteed Minimum Compensation\\
$k$ &Ratio of the cumulative penalty locked by payer \\
&and the payment value locked by payee\\
$\gamma^{\zeta,k}$ &Rate of griefing-penalty in $\textrm{HTLC-GP}^{\zeta}$ for a given $\zeta$ and $k$\\
  $\tilde{n}^{\zeta,k}$ & Maximum path length in $\textrm{HTLC-GP}^{\zeta}$ for a given $\zeta$ and $k$, $\tilde{n}^{\zeta,k} \leq n$\\

\hline
\end{tabular}
}
\end{center}
\caption{\textcolor{black}{Notations used in the paper}}
\label{tab1}
\end{table}

\textcolor{black}{Given an instance of payment in $G$, where a payer $S$ wants to transfer $\alpha$ coins to a payee $R$ where $S, R \in V$. \textcolor{black}{The notations used in defining the model is summarized in Table \ref{tab1}.} We discuss how the payment is routed in the network:\\
(i) $S$ finds a path $P$ of length $\kappa: \kappa \in \mathbb{N}$ that connects it to $R$. The maximum allowed path length for routing is $n$ so $|P|=\kappa \leq n$. We denote $P=\langle U_0\rightarrow U_1 \rightarrow U_2 \ldots \rightarrow U_{\kappa} \rangle$, where $U_0=S$ and $U_{\kappa}=R$ and $locked(U_{i-1},U_i)>\alpha, \forall (U_{i-1},U_{i}) \in E, i \in [1,\kappa]$. \\
(ii) Criteria for a node to route the payment: A node $U_{i-1}$ can lock $\alpha_{i-1}$ coins in an off-chain \emph{HTLC} formed with $U_i$ if $remain(U_{i-1},U_{i})\geq \alpha_{i-1} : \alpha_{i-1}=\alpha_0 - \Sigma_{k=1}^{i-1} f(\alpha_{i-1}), i \in [1,n]$. Node $U_{i-1}$ gets a processing fee $f(\alpha_{i-1})$. If $U_i$ claims the coins, the residual capacity is updated as follows : $remain(U_{i-1},U_{i})=remain(U_{i-1},U_{i})-\alpha_{i-1}$ and $remain(U_{i},U_{i-1})=remain(U_{i},U_{i-1})+\alpha_{i-1}$. \\
(iii) Once the path is decided, $U_{\kappa}$ generates a payment condition $H=\mathcal{H}(x)$ and shares it with $U_0$. The latter forwards the payment across $P$ by forwarding \emph{HTLC}'s. The \emph{HTLC} timeout period between any pair $U_{i-1}$ and $U_{i}$ is set to $t_{i-1}, i \in [1,\kappa]$. If $U_i$ chooses to resolve the \emph{HTLC} just before timeout period, it responds at time $t_{i-1}-\delta$, where $\delta \rightarrow 0$. The least \emph{HTLC} timeout period $D$ is assigned for the contract between $U_{\kappa-1}$ and $U_{\kappa}$, i.e., $t_{\kappa-1}=D$.\\
(iv) Lightning Network uses the Sphinx protocol \cite{danezis2009sphinx} while forwarding \emph{HTLC}s. It is a form of onion routing where none of the intermediate parties have any information regarding the routing path except their immediate neighbors. Thus, a node $U_{i}$ upon receving a request, knows that a conditional payment request came from $U_{i-1}$ and it must be forwarded to node $U_{i+1}$ where $i \in [1,\kappa-1]$.} 

\emph{System Assumption}: We discuss some of the assumptions:\\
(i) All the nodes in the network are rational \cite{azouvi2019sok}, \cite{garay2013rational}\footnote{For our model, we rule out any altruistic and Byzantine behavior and focus on the most typical scenario where participants are rational. However, the Lightning network may have Byzantine as well as altruistic nodes. We leave the analysis of griefing attack in the BAR model \cite{aiyer2005bar} as future work.}. Rational processes always seek to maximize their expected utility. They may deviate or not choose to participate in a protocol depending on the situation. \\
(ii) We assume that a channel between $U_{i-1}$ and $U_{i}$ is unilaterally funded by $U_{i-1}, i \in [1,\kappa]$, i.e., $locked(U_{i},U_{i-1})=0$. \\
(iii) We define a function $O: \mathbb{R}^+ \times \mathbb{W} \times \mathbb{R}^+ \rightarrow \mathbb{R}^+ \cup \{0\}$, where $O(r_U,t,val)$ is the expected revenue a node $U$ would have earned had it utilized the amount $val$ for processing transactions in a period of $t$ units given that $r_U$ is the rate of payments processed by $U$ per unit time. In other words, $O$ defines the \emph{opportunity cost} \cite{buchanan1991opportunity}. 

The primary source of revenue for a routing node in the Lightning Network is the fee obtained by processing transactions \cite{osuntokun2020lightning}. Also, the arrival of payment in a channel follows a Poisson process \footnote{All the papers assumed arrival of transaction in Bitcoin blockchain as a Poisson process but the validity of the assumption was not verified. It was first analyzed in \cite{gebraselase2021transaction,gebraselase2021analysis} and the authors have reported that transactions’ inter-arrival times can be approximately fitted with an exponential distribution, which partially supports the Poisson arrival assumption but with noticeable deviation. In our paper, we consider the arrival of transaction in Bitcoin blockchain following a Poisson process and the same holds true for transactions arriving in Lightning Network.} \cite{gebraselase2021transaction}, \cite{guasoni2021lightning}, \cite{kawase2017transaction}. $U$ expects each transaction size to be $per\_tx\_val$. Given the number of coins locked is $val$, the number of transactions $U$ expects to receive in period of $t$ units is $J=\frac{val}{per\_tx\_val}$. Given $X$ is the number of transactions in that interval or $X \sim Poisson(r_Ut)$, we have. 
\begin{equation}
\begin{matrix}
P(X=x)=\frac{e^{-r_Ut}(r_Ut)^{x}}{x!}\\
\end{matrix}
\end{equation}
where $0\leq x \leq J$. Expected number of transactions in $t$ unit of time
\begin{equation}
\begin{matrix}
E(X)=\sum\limits_{x=0}^{J} xP(X=x)
\end{matrix}
\end{equation}
The fee earned by processing a transaction of size $per\_tx\_val$ is defined as \cite{fee1},\cite{fee2}:
\begin{equation}
Fee_{per\_tx\_val}=base\_fee+fee\_rate\times per\_tx\_val
\end{equation}
Thus, the revenue $U$ expects to earn within period $t$ or in other words, the opportunity cost $O(r_U,t,val)$ is 
\begin{equation}
\begin{matrix}
O(r_U,t,val)=E(X) Fee_{per\_tx\_val}
\end{matrix}
\end{equation}




\subsection{Attacker Model}
An attacker with budget $\mathcal{B}_{EX}$ tries to disrupt the network by incentivizing a certain fraction of nodes to mount the griefing attack \footnote{It is a standard practice to consider external incentives and several works have adhered to this model \cite{garay2013rational}, \cite{azouvi2019sok}.}. We make the following assumptions - (i) if a node has accepted the bribe, then it implies that the expected earning by cooperating with other nodes is lower than the bribe, and hence it has chosen to be a corrupt node. Such nodes act as per the instructions received from the attacker, and (ii) a corrupt node knows whether another node is corrupt or uncorrupt. An uncorrupt nodes lacks this information. 

We discuss the bribe offered to a corrupt node and the method for mounting a griefing attack:\\
\textcolor{black}{(a) \textit{Bribe offered per attack}. Given a payment send across the network is of value $\alpha$, the attacker fixes the bribe offered to a node to $L$ coins, where $L= \alpha+I_{D,\alpha}+C$. Here $C$ is the auxiliary cost for routing payment and opening new channels, if needed. $I_{D,\alpha}$ coins are used to compensate the node for keeping $\alpha$ coins unutilized for the next $D$ units of time. We assume $I_{D,\alpha}\approx 2 O(r_{U},D,\alpha), \forall U\in V$ so that a corrupt node gains at least $O(r_{U},D,\alpha)$ inspite of locking $\alpha$ coins.\\ 
(b) \textit{Method for mounting Griefing Attack.} The attacker instructs the corrupt node to execute a self-payment (i.e., $S=R$) of $\alpha$ coins via a route of maximum allowed path length $n$ in order to maximize the damage. The least HTLC timeout period is $t_{n-1}\approx D$. After the conditional payment reaches the payee $U_n$, it intentionally stops responding, locking a collateral of $(n-1)\alpha$ for the next $D$ units in the path routing payment.}

\subsection{Game Model}

\textcolor{black}{(i) \emph{Choice of players}: We assume that all the miners in the underlying blockchain are honest, and only nodes in Lightning Network can be the strategic players. In the path $P$, a node $U_{i-1}$ locks an amount $\alpha_{i-1}$ with the off-chain contract formed with $U_{i}$, hence it will be bothered with $U_{i}'s$ nature and the corresponding action. Except for $U_0$, none of the intermediaries routing the payment knows the recipient's identity. $U_{i-1}$ makes a decision of whether to forward a payment to $U_i$ based on the belief of the $U_i's$ type (discussed next). We model the griefing attack as an interaction between pair of nodes $U_{i-1}$ and $U_{i}, i \in [1,\kappa]$ in path $P$ routing payment in the network.  Player forms a belief about the type of the other players based on either their position in the network or past interaction.} 

(ii) \emph{Action Space}: We define the actions of $U_{i-1}$ and $U_i$:

\begin{itemize}
\item \emph{$U_{i-1}$'s action ($S_{U_{i-1}}$)}: It can either \textit{forward (F)} the conditional payment to $U_{i}$ or it can choose to \emph{not forward (NF)}. If it chooses to forward the payment, it forms a contract with $U_{i}$, locking the designated amount in the channel $id_{i-1,i}$ for time $t_{i-1}$, which is the \emph{HTLC} timeout period. If $i>1$, then $U_{i-1}$ gets a fee of $f(\alpha_{i-1})$ from $U_{i-2}$ contingent to the release of solution by $U_i$. For $U_0$, the satisfaction level is proportional to success of payment. In this case, we consider $f(\alpha_0)$ as the level of satisfaction. If $U_i$ delays, then opportunity cost of coins locked in the off-chain contract increases, and a loss is incurred. If $U_{i}$ doesn't respond within $t_{i-1}$, then $U_{i-1}$ closes the channel and withdraws its coins from the contract.

\item \emph{$U_i$'s action ($S_{U_{i}}$)}: If $U_{i-1}$ has forwarded the payment, then $U_{i}$ can choose its action from the following: \emph{accept the payment} or \emph{Ac}, \emph{reject the payment} or \emph{Rt}, \emph{wait and then accept} or \emph{W \& Ac}, \emph{wait and then reject} or \emph{W \& Rt}, and \emph{grief} or \emph{Gr}. If $U_{i-1}$ does not forward the payment, the game aborts.


\end{itemize} 

We define the sequential \emph{Bayesian} game $\Gamma_{HTLC}$ as the tuple $\Gamma_{HTLC}=\langle N, (\Theta_{U_{i-1}},\Theta_{U_{i}}), (S_{U_{i-1}},S_{U_{i}}),p_{U_{i-1}},(u_{U_{i-1}},u_{U_{i}})\rangle$ \cite{narahari2012game}, where $N =\{U_{i-1},U_{i}\}$ where $i \in [1,\kappa]$. The type of player $U_{i}$ is defined as $\Theta_{U_{i}}=\{\textrm{Corrupt(co),} \textrm{ Uncorrupt (uco)}\}$.
The probability function $p_{U_{i-1}}$ is a function from $\Theta_{U_{i-1}}$ into $p(\Theta_{U_{i}})$, where the $p(\Theta_{U_{i}})$ denotes the set of probability distribution over $\Theta_{U_{i}}$, i.e., $p_{U_{i-1}}(Corrupt)=\theta_i$, $p_{U_{i-1}}(Uncorrupt)=1-\theta_i$. The payoff function $u_{k}: \Theta \times S \rightarrow \mathbb{R}$ for any player $k \in \{U_{i-1},U_{i}\}$, where $\Theta=\Theta_{U_{i}}$ and $S=S_{U_{i-1}}\times S_{U_{i}}$, is such that for any profile of actions and any profile of types $(\hat{\theta},s) \in \Theta \times S$, specifies the payoff the player $k$ would get, if the player's actual type were all as in $\hat{\theta}$ and the players all chose their action as in $s$.

%

\subsubsection{Preference Structure}
  \begin{figure}[htbp]
\hspace*{-1.3cm}
\raggedright
\begin{minipage}{9.5cm}
    \includegraphics[width=14.5cm,height=4.5cm,left]{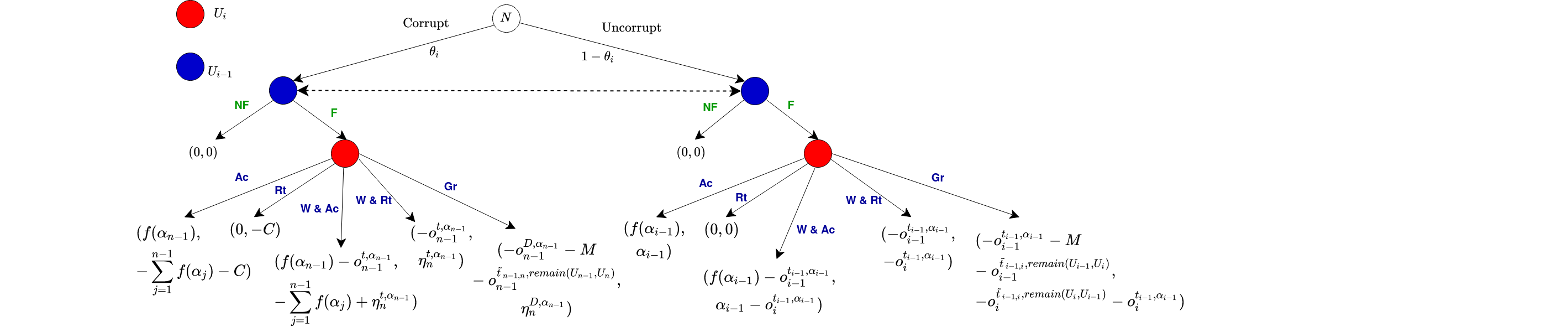}
    \caption{Extensive form of game $\Gamma_{HTLC}$ }
    \label{ext1}
\end{minipage}
\end{figure}

The game begins with Nature (\textbf{N}) choosing the type of $U_{i}$, either \emph{corrupt} or \emph{uncorrupt}, respectively. $U_{i-1}$ believes that a corrupt $U_{i}$ will be selected with probability $\theta_i$, whereas an uncorrupt $U_{i}$ will be selected with probability $1-\theta_i$. After \textbf{N} makes its move, $U_{i-1}$ selects its strategy based on the belief of $U_i$'s type. $U_i$ chooses its strategy after $U_{i-1}$ has forwarded the payment. The extensive form is represented in Fig.\ref{ext1}. If $U_{i-1}$ chooses not to forward, then either party receives a payoff 0 since no off-chain contract got established, i.e., $u_{U_{i-1}}(\theta_b,NF,s_b)=u_{U_{i}}(\theta_b,NF,s_b)=0,  \theta_b \in \Theta_{U_{i}}$ and $s_b \in S_{U_{i}}$. \textcolor{black}{We analyze the payoff of $U_{i-1}$ when it has chosen $F$}:
\begin{enumerate}

\itemsep0em 

\item[\textbf{A}.] \textcolor{black}{If \textbf{N} had chosen an \emph{uncorrupt} $U_{i}$, then the payoffs are defined as follows:}

\textcolor{black}{(a) Instantaneous Response, i.e., $t\rightarrow 0$: If \emph{$U_{i}$ accepts the payment}, then $U_{i-1}$ gets processing fee $f(\alpha_{i-1})$ from its preceding neighbour $U_{i-2}$ and $U_i$ gets $\alpha_{i-1}$ coins from $U_{i-1}$. If \emph{$U_{i}$ rejects the payment} then none of them gains anything, i.e, $u_{U_{i-1}}(uco,F,Rt)= u_{U_{i}}(uco,F,Rt)=0$.} \\

\textcolor{black}{(b) Delayed Response, i.e., $0<t < t_{i-1}$: If \emph{$U_{i}$ waits and then accepts the payment after $t$ units} then $u_{U_{i-1}}(uco,F,W \ \& \ Ac)=f(\alpha_{i-1})-o_{i-1}^{t,\alpha_{i-1}}$. $U_{i-1}$ can earn $f(\alpha_{i-1})$ only after $U_i$ resolves the payment but it suffers a loss due to $\alpha_{i-1}$ coins remaining locked in channel $id_{i-1,i}$. $O(r_{U_{i-1}},t,\alpha_{i-1})$ is the opportunity cost of locked coins, also denoted as $o_{i-1}^{t,\alpha_{i-1}}$. Simultaneoulsy, $U_{i}$ loses the opportunity to earn profit by utilizing $\alpha_{i-1}$ coins for the next $t$ unit of time. The expected profit that $U_i$ could have made using $\alpha_{i-1}$ within the next $t$ units is $O(r_{U_i},t,\alpha_{i-1})$, also denoted as $o_i^{t,\alpha_{i-1}}$. Thus the payoff $u_{U_{i}}(uco,F,W \ \& \ Ac)=\alpha_{i-1}-o_i^{t,\alpha_{i-1}}$.}

\textcolor{black}{If \emph{$U_{i}$ waits and then rejects the payment after time $t$ units} then the payoff of $U_{i-1}$, $u_{U_{i-1}}(uco,F,W \ \& \ Rt)=-o_{i-1}^{t,\alpha_{i-1}}$, and payoff of $U_{i}$, $u_{U_{i}}(uco,F,W \ \& \ Rt)=-o_{i}^{t,\alpha_{i-1}}$.} \\

\textcolor{black}{(c) \emph{$U_{i}$ griefs}: If $U_{i}$ fails to respond within time $t_i$, $U_{i-1}$ will close the channel by going on-chain. $U_{i-1}$ cannot utilize $\alpha_{i-1}$ coins locked in the off-chain contract. The opportunity cost is $O(r_{U_{i-1}},t_{i-1}, \alpha_{i-1})$, also denoted as $o_{i-1}^{t_{i-1},\alpha_{i-1}}$. Additionally, due to closure of channel, $U_{i-1}$ fails to utilize the residual capacity $remain(U_{i-1},U_i)$ for the next $T-(t_{i-1}+t_{contract\_initiate}-t_{i-1,i})$ unit of time, where $t_{i-1,i}$ is the timestamp at which channel $id_{i-1,i}$ was opened and $t_{contract\_initiate}$ is the current timestamp at which the off-chain contract got initiated in the channel. We use a shorter notation $\tilde{t}_{i-1,i}$ to denote $T-(t_{i-1}+t_{contract\_initiate}-t_{i-1,i})$. The opportunity cost of the remaining balance is  $O(r_{U_{i-1}},T-(t_{i-1}+t_{contract\_initiate}-t_{i-1,i}),remain(U_{i-1},U_i))$ or $o_{i-1}^{\tilde{t}_{i-1,i},remain(U_{i-1},U_i)}$. Along with that $U_{i-1}$ has to pay the transaction fee $M$ for settling on-chain. Hence, payoff $u_{U_{i-1}}(uco,F,Gr)=-o_{i-1}^{t_{i-1},\alpha_{i-1}}-o_{n-1}^{\tilde{t}_{i-1,i},remain(U_{i-1},U_i)}-M$.}

\textcolor{black}{If $U_{i-1}$ had previously transferred coins to $U_{i}$ then $remain(U_i,U_{i-1})>0$. In that case, $U_i$ incurs a loss $O(r_{U_{i}},T-(t_{i-1}+t_{contract\_initiate}-t_{i-1,i}),remain(U_{i},U_i-1))$, also denoted as $o_{i}^{\tilde{t}_{i-1,i},remain(U_{i},U_{i-1})}$, due to closure of channel after timeout period $t_{i-1}$. Additionally, it loses $o_i^{t_{i-1},\alpha_{i-1}}$, as it fails to earn and utilize the coins for other purpose. Hence, payoff $u_{U_{i}}(uco,F,Gr)=-o_{i}^{\tilde{t}_{i-1,i},remain(U_{i},U_{i-1})}-o_i^{t_{i-1},\alpha_{i-1}}$. Since $U_i$ doesn't go on-chain for settling the transaction, we do not subtract $M$ from the payoff.}


\item[\textbf{B}.] \textcolor{black}{If \textbf{N} had chosen an \emph{corrupt} node, then the latter executes a self-payment of amount $\alpha$ via a path of length $n$. Thus we analyze this as a game between $U_{n-1}$ and $U_n$, where the latter is the corrupt node. The amount forwarded is $\alpha+\sum\limits_{i=1}^{n-1}f(\alpha_{i})$, where, $f(\alpha_{i})$ is the fee charged by an intermediate node $U_i$. Since the cost incurred per payment is $C$ and the corrupt node has to keep $\alpha$ coins locked for time $t_{n-1}=D$, the bribe offered must compensate for all these costs. The amount of bribe offered by the attacker is $L$ where $L = \alpha+C+I_{D,\alpha}$, where $I_{D,\alpha}\approx 2 o_n^{D,\alpha}$. Since the purpose of $U_n$ is to mount attack, it would not be interested in performing payments like other participants. This implies that $U_{n}$ has not accepted any payment from $U_{n-1}$ and $remain(U_n,U_{n-1})=0$. We analyze each case as follows:}

\textcolor{black}{(a) Instantaneous Response, i.e., $t \rightarrow 0$: If \emph{$U_n$ accepts the payment} then it ends up losing approximatey $\sum\limits_{i=1}^{n-1}f(\alpha_{i})$, as it needs to pay $(n-1)$ intermediaries. It had already incurred a cost $C$. $U_{n-1}$ has successfully forwarded the amount. Thus, payoffs are $u_{U_{n-1}}(co,F,Ac)=f(\alpha_{n-1})$ and $u_{U_{n}}(co,F,Ac)=-C-\sum\limits_{i=1}^{n-1}f(\alpha_{i})$. If \emph{$U_i$ or $U_{n}$ rejects the payment} then $u_{U_{n-1}}(co,F,Rt)=0$ and $u_{U_{n}}(co,F,Rt)=-C$.}\\


\textcolor{black}{(b) Delayed Response, i.e., $0<t\leq D-\delta$: If \emph{$U_{n}$ waits and then accepts the payment after $t$ units} then payoff of $U_{n-1}$ is same as the payoff it had obtained when $U_{n}$ is not corrupt and chooses to wait and accept the payment. Thus $u_{U_{n-1}}(co,F,W \ \& \ Ac)=f(\alpha_{n-1})-o_{n-1}^{t,\alpha_{n-1}}$. $\eta_n^{t,\alpha_{n-1}}$ defines the net profit received by $U_n$ for keeping $\alpha_{n-1}$ coins unutilized till time $t$, where:
\begin{equation}
\label{lock}
\eta_n^{t,\alpha_{n-1}}=
    \begin{cases}
    -C-O(r_{U_{n}},t,\alpha_{n-1}),  0<t<D-\delta\\
    L-C-O(r_{U_{n}},D-\delta,\alpha_{n-1}), \\  \textrm{otherwise}
    \end{cases}
\end{equation}
If $U_n$ delays till time $t<D-\delta$, it loses the setup cost and the revenue had it utilized $\alpha_{n-1}$ for $t$ units of time. If $U_n$ delays for at least $D-\delta$, it gets paid for the work done, i.e. $\eta_n^{D-\delta,\alpha_{n-1}}$. Since $\delta \rightarrow 0$, $\eta_n^{D-\delta,\alpha_{n-1}} \approx \eta_n^{D,\alpha_{n-1}}= L-C-O(r_{U_{n}},D,\alpha_{n-1})$. But $I_{D,\alpha_{n-1}}\approx  2 o_{n}^{D,\alpha_{n-1}}$, which implies $L-C-o_n^{D,\alpha_{n-1}} =\alpha+C+I_{D,\alpha_{n-1}}-C-o_n^{D,\alpha_{n-1}} \approx \alpha+o_n^{D,\alpha_{n-1}}$. Upon accepting a self-payment, it ends up paying a processing fee to $n-1$ intermediaries. Thus, the payoff of $U_{n}$, $u_{U_{n}}(co,F,W \ \& \ Ac)=-\sum\limits_{i=1}^{n-1}f(\alpha_{i})+\eta_n^{t,\alpha_{n-1}}$.}

\textcolor{black}{If \emph{$U_{n}$ waits and then rejects the payment after $t$ units} then $u_{U_{n-1}}(co,F,W \ \& \ Rt)=-o_{n-1}^{t,\alpha_{n-1}}$ and $u_{U_{n}}(co,F,W \ \& \ Rt)=\eta_n^{t,\alpha_{n-1}}$.}\\

\textcolor{black}{(c) \emph{$U_{n}$ griefs}: It gets an incentive $L$ coins from attacker. The loss is summation of $C$, which is the cost for mounting the attack, and the opportunity cost $o_n^{D,\alpha_{n-1}}$. Since the channel is unilaterally funded by $U_{n-1}$, $remain(U_n,U_{n-1})=0$. Thus there is no loss associated due to closure of channel. The payoff of $U_{n-1}$ is the same as the payoff it had obtained when $U_{n}$ is uncorrupt and chooses to grief. Thus, $u_{U_{n-1}}(co,F,Gr)=-o_{n-1}^{D,\alpha_{n-1}}-o_{n-1}^{\tilde{t}_{n-1,n},remain(U_{n-1},U_n)}-M$ and $u_{U_{n}}(co,F,Gr)=L-C-o_n^{D,\alpha_{n-1}}=\eta_{n}^{D,\alpha_{n-1}}$.}


\end{enumerate}

\subsubsection{Game Analysis}
\label{ga1}

\textcolor{black}{We infer from the payoff model that the corrupt node can select either of the strategies for mounting the attack - (i) \emph{Reject the payment just before lock time $D$ elapses}: $U_n$ rejects the conditional payment forwarded by $U_{n-1}$ off-chain just before the contract's lock time elapses. (ii) \emph{$U_n$ does not respond}: This is as per the conventional definition of griefing. $U_{n-1}$ closes the channel unilaterally after the contract's lock time expires.}


\textcolor{black}{The expected payoff of $U_{i-1}$ is calculated by applying backward induction on the game tree $\Gamma_{HTLC}$. If $U_{i-1}$ plays \emph{F}; an \emph{uncorrupt} $U_{i}$ chooses \emph{Ac} as its best response since $u_{U_i}(uco,F, Ac)\geq u_{U_i}(uco,F,s')$, $\forall s' \in S_{U_i}$; a corrupt node (also $U_n$) can choose either to \emph{grief} or \emph{Wait \& Reject} at $D-\delta$ as its best response since $u_{U_n}(co,F,Gr)=u_{U_n}(co,F, W \ \& \ Rt \textrm{ at time }D-\delta)\geq u_{U_n}(co,F,s''), \forall s'' \in S_{U_n}$. A corrupt node applies mixed strategy, either choosing to \emph{Grief} with probability $1-q$ or it can \emph{Wait and Reject} at time $D-\delta$ with probability $q$. The expected payoff of $U_{i-1}$ for selecting \emph{F}, denoted as $\mathbb{E}_{U_{i-1}}(F)$, is $\theta_i \Big(-q o^{D-\delta,\alpha_{n-1}}_{n-1}+(1-q)( -o_{n-1}^{D,\alpha_{n-1}}-o_{n-1}^{\tilde{t}_{n-1,n},remain(U_{n-1},U_n)}-M) \Big)+ (1-\theta_i)f(\alpha_{i-1})$ and expected payoff for selecting \emph{NF}, denoted as $\mathbb{E}_{U_{i-1}}(NF)$, is 0.}

\textcolor{black}{Since $\delta \rightarrow 0$, we consider $o^{D-\delta,\alpha_{n-1}}_{n-1}\approx o^{D,\alpha_{n-1}}_{n-1}$. If $\mathbb{E}_{U_{i-1}}(F)>\mathbb{E}_{U_{i-1}}(NF)$ then $U_{i-1}'s$ best response is \emph{F} else it chooses \emph{NF}. We derive that $U_{i-1}$ chooses \emph{F} if $\theta_i<\frac{f(\alpha_{i-1})}{o^{D,\alpha_{n-1}}_{n-1}+f(\alpha_{i-1})+(1-q)\big(o_{n-1}^{\tilde{t}_{n-1,n},remain(U_{n-1},U_n)}+M\big)}$, else it chooses \emph{NF}; \emph{corrupt}  $U_{i}$ can either choose \emph{Grief} or \emph{Wait \& Reject} at time $D-\delta$; \emph{uncorrupt} $U_{i}$ chooses \emph{Accept}; is a perfect Bayesian equilibrium.}

%
%


\section{Hashed Timelock Contract with Griefing Penalty or HTLC-GP}
\label{countermeasure}
The protocol Hashed Timelock Contract with Griefing Penalty or \emph{HTLC-GP} \cite{agrief} has been developed on top of \emph{HTLC}. The concept of the griefing penalty is used in the protocol for countering the griefing attack. If the party griefs, it gets penalized, and the amount locked is distributed as compensation amongst the affected parties. The total penalty charged is proportional to the summation of the collateral locked in each off-chain contract instantiated on the channels routing payment. \emph{Collateral locked} in an off-chain contract is the product of coins locked in the off-chain contract and timeout period.

\begin{figure}[!ht]
    \centering
    \includegraphics[width=8cm]{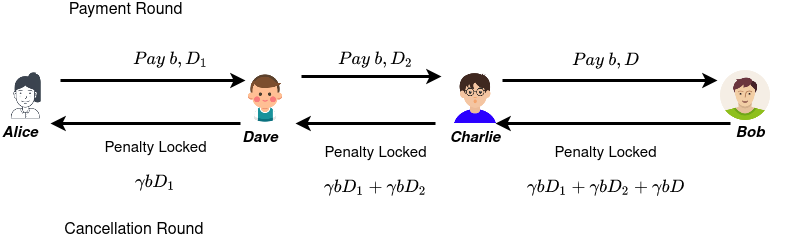}
    \caption{Formation of contract in \textit{HTLC-CP}}
    \label{fig12}
\end{figure}

\emph{HTLC-GP} is a two-round protocol where the first round, or \emph{Cancellation round}, involves locking of penalty. The round is initiated by the payee and proceeds in the reverse direction. The penalty locked by a party serves as a guarantee against a payment forwarded. The second round, or the \emph{Payment round}, involves locking the payment value in off-chain contracts from payer to payee. We explain the protocol with the help of an example shown in Fig. \ref{fig12}. \emph{Alice} wants to transfer $b$ units to \emph{Bob}. Each party that forwards a payment must be guaranteed by its counterparty to receive compensation if there is an incidence of a griefing attack. The compensation charged must be proportional to the collateral locked in the path. We define this proportionality constant as the \emph{rate of griefing-penalty per unit time}, denoted as $\gamma$. The first round termed as \emph{Cancellation round} proceeds in the following way: \emph{Bob} has to lock $\gamma b D_1+\gamma b D_2+\gamma b D$ coins for duration $D$. This amount is the cumulative penalty to be distributed among \emph{Alice, Dave} and \emph{Charlie}, if Bob griefs. After \emph{Charlie} receives the cancellation contract, he locks $\gamma b D_1+\gamma b D_2$ in the contract formed with \emph{Dave} for $D_2$ units. The latter locks $\gamma b D_1$ coins in the contract formed with \emph{Alice} for $D_1$ units of time. The second round, termed as \emph{Payment round} proceeds similarly as in \emph{HTLC}. Payment value $b$ is forwarded from \emph{Alice} to \emph{Bob} via the intermediaries. Since the least timeout period is $D$, one might question why \emph{Bob} must take into account the lock time of the other contracts while locking penalty. If \emph{Bob} locks $3\gamma b D$ coins, \emph{Charlie} locks $2 \gamma b D_2$  and \emph{Dave} locks $\gamma b D_1$ as penalty, and \emph{Bob} griefs, then \emph{Charlie} keeps the compensation $\gamma b D$ and forwards $2\gamma b D$ to \emph{Dave}. \emph{Dave} is greedy and refuses to cancel the off-chain contract with \emph{Charlie}. After elapse of $D_2$, he goes on-chain and claims $2\gamma D_2$. Since $D_2>D$, \emph{Charlie} incurs a loss of $2\gamma b (D_2-D)$. Thus, we account for the lock time of each contract while calculating compensation to prevent the loss of coins of uncorrupt parties.

Suppose \emph{Bob} griefs. He has to pay a compensation of $\gamma b D_1+\gamma b D_2+\gamma b D$ units to \emph{Charlie}, as per the terms of the contract. After $D$ expires, \emph{Charlie} goes on-chain. He closes the channel, unlocks $b$ coins, and claims compensation. He requests \emph{Dave} to cancel the off-chain contract, offering a compensation of $\gamma b D_1 +\gamma b D_2$. \emph{Dave} cancels the contract off-chain, unlocks $b$ units from the contract, and claims the compensation from \emph{Charlie}. If \emph{Charlie} decides to grief, \emph{Dave} can claim the compensation by going on-chain and closing the channel. \emph{Dave} requests \emph{Alice} to cancel the contract by offering a compensation of $\gamma  b D_1$. Except for \emph{Bob}, none of the parties lose coins. In the next section, we formulate a game model for griefing attacks in \emph{HTLC-GP} and study its effectiveness.


\section{Analysis of griefing attack in HTLC-GP}
\label{gpanalysis}
\subsection{System Model}
\textcolor{black}{For payment of $\alpha$ from $S$ to $R$ via $(\kappa-1)$ intermediaries, where $\kappa\in \mathbb{N}, \kappa\leq n$, we denote the cumulative compensation to be locked by $U_{i}$ if $U_{i-1}$ forwards $\alpha_{i-1}$ as $Z_{\alpha_{i-1},i}, i \in [1,\kappa]$ where $Z_{\alpha_{i-1},i}=Z_{\alpha_{i-2},i-1}+c_{\alpha_{i-1},i-1}$; $c_{\alpha_{i-1},i-1}$ is the compensation charged by node $U_{i-1}$ and $Z_{\alpha_{i-2},{i-1}}$ is used for compensating other nodes $U_j, j<i$. Note that $Z_{val,0}=0, \forall val \in \mathbb{R}^{+}, \alpha_{\kappa-1}=\alpha$. $c_{\alpha_{i-1},i-1}$ charged by a node $U_{i-1}$, must be proportional to the collateral it has locked in the off-chain contract formed with node $U_{i}$ for timeout period $t_{i-1}, i \in [1,\kappa]$. $t_{i-1}=t_{i}+\Delta$, $t_{\kappa-1}=D$ and $c_{\alpha_{i-1},i-1}=\gamma \alpha_{i-1}t_{i-1}$, $\gamma$ being the rate of griefing-penalty per unit time. An off-chain contract between node $U_{i-1}$ and $U_{i}$ requires $U_{i-1}$ locking an $\alpha_{i-1}$ coins and $U_{i}$ must lock $Z_{\alpha_{i-1},i}$ coins. Here $Z_{\alpha_{\kappa-1},\kappa}$ denoted as $Z_{\alpha}=\sum\limits_{i=0}^{\kappa-1} c_{\alpha_i,i}$ is locked by $U_{\kappa}$ as guaranteed compensation against the amount locked by party $U_{\kappa-1}$. Again, $U_{\kappa-1}$ has locked $Z_{\alpha_{\kappa-2},\kappa-1}$ with the contract formed with $U_{\kappa-2}$ for time $t_{\kappa-1}$.}

\emph{Change in System Assumption}: The assumptions taken here is same as in Section \ref{sys-assumption1}, except that the payment channel is considered to be dual-funded. This implies that in channel $id_{i-1,i},i\in [1,\kappa]$, both the parties $U_{i-1}$ and $U_{i}$ lock coins i.e., $locked(U_{i-1},U_{i})>0$ and $locked(U_{i},U_{i-1})>0$.

\subsection{Attacker Model}
\label{at1}
\textcolor{black}{If a corrupt node routes a self-payment via maximum allowed path length $n$, then the recipient (i.e., $U_n$) has to lock extra coins as guarantee. Cost of the attack increases. However, the attacker does not increase the incentive offered per attack. Thus, $U_{n}$ is forced to distribute $\alpha$ coins between the cumulative penalty locked in the contract formed with $U_{n-1}$ and the amount to be forwarded for payment. This implies $\alpha$ is the summation of transaction value $v+\sum\limits_{i=1}^{n-1}f(v_i)$ and the cumulative penalty $Z_v=\sum\limits_{i=0}^{n-1}c_{v_i,i}=\gamma  \sum\limits_{j=0}^{n-1} v_j t_j$ where $v_j=v_0-\sum\limits_{k=1}^{j}, j \in [0,n-1]$. Since $\sum\limits_{i=1}^{n-1}f(v_i)<<v$, we consider $v_0+Z_v \approx v+Z_v=\alpha$ or, $v=\frac{\alpha}{1+\gamma \sum\limits_{j=0}^{n-1} t_j}$. $U_n$ executes a self-payment of $v$ coins.} 


\subsection{Game Model}
\begin{figure}[!ht]
  \hspace*{-0.2cm}
    \centering
    \includegraphics[width=	10cm]{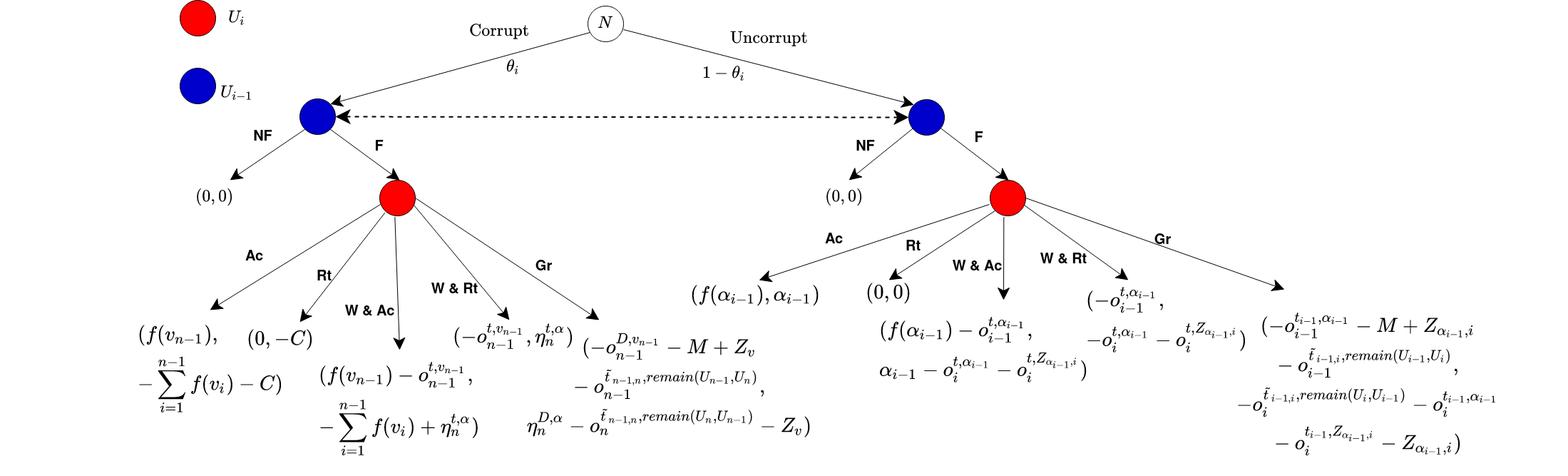}
    \caption{Extensive form of the game $\Gamma_{HTLC-GP}$}
    \label{ext2}
\end{figure}

Given a payment is routed via a path of length $\kappa$ using \emph{HTLC-GP}, $U_{\kappa}$ locks penalty in the contract formed with $U_{\kappa-1}$ in the first round locking phase. However, the former has the power to unlock the coins anytime by releasing the preimage of the cancellation hash. $U_{\kappa-1}$ will accept the contract if it thinks that the expected payoff upon forwarding the payment contract in the second round will be greater than the expected payoff on not forwarding the same. Only then it will lock the penalty in the off-chain contract with $U_{\kappa-2}$. This holds for any pair $U_{i-1}$ and $U_i$ in path $P$. If $U_{i-1}$ accepts to form a contract with $U_i$ in the first round of \emph{HTLC-GP}, it implies that it will forward the conditional payment to $U_i$ in the second round as well. Thus we merge both the first and second round while studying the interaction between any two parties $U_{i-1}$ and $U_i$. We analyze the payoff assuming both parties lock their coins in a single off-chain contract instead of two separate contracts. We adapt the model $\Gamma_{HTLC}$ and propose the two-party game model $\Gamma_{HTLC-GP}$ for griefing attack in \emph{HTLC-GP}. The extensive form of the game $\Gamma_{HTLC-GP}$ is shown in Fig.\ref{ext2}

\subsubsection{Preference Structure}
When $U_{i-1}$ chooses not to forward, both $U_{i-1}$ and $U_i$ receives a payoff 0 since no off-chain contract got established, i.e., $u_{U_{i-1}}(\theta_b,NF,s_b)=u_{U_{i}}(\theta_b,NF,s_b)=0,  \theta_b \in \Theta_{U_{i}}$ and $s_b \in S_{U_{i}}$. We analyze the payoff of each case when $U_{i-1}$ chooses to forward:
\begin{enumerate}
\itemsep0em 

\item[\textbf{A}.] If \textbf{N} had chosen an \emph{uncorrupt} $U_{i}$, then the payoffs are defined as follows:

(a) Instantaneous Response, i.e., $t\rightarrow 0$: Upon instant acceptance or rejection of payment, the payoffs are the same as that observed in $\Gamma_{HTLC}$.\\

(b) Delayed Response, i.e., $0<t < t_i$: If \emph{$U_{i}$ waits and then accepts the payment after $t$ units} then $u_{U_{i-1}}(uco,F,W \ \& \ Ac)=f(\alpha_{i-1})-o_{i-1}^{t,\alpha_{i-1}}$. $U_{i}$ has to keep $Z_{\alpha_{i-1},i}$ coins locked in the contract established in channel $id_{i-1,i}$. Thus, it faces loss not only due to delay in claiming $\alpha_{i-1}$ coins from $U_{i-1}$ but also due to unutilization of $Z_{\alpha_{i-1},i}$. The expected profit that could have been made using $\alpha_{i-1}$ and $Z_{\alpha_{i-1},i}$ within the next $t$ units is $O(r_{U_{i}},t,\alpha_{i-1})$, also denoted as $o_{i}^{t,\alpha_{i-1}}$, and $O(r_{U_i},t,Z_{\alpha_{i-1},i})$, denoted as $o_i^{t,Z_{\alpha_{i-1},i}}$. Thus, $u_{U_{i}}(uco,F,W \ \& \ Ac)=\alpha_{i-1}-o_i^{t,\alpha_{i-1}}-o_i^{t,Z_{\alpha_{i-1},i}}$.

If \emph{$U_{i}$ waits and then rejects the payment after $t$ units} then the payoff of $U_{i-1}$ is $u_{U_{i-1}}(uco,F,W \ \& \ Rt)=-o_{i-1}^{t,\alpha_{i-1}}$ and payoff of $U_{i}$ is $u_{U_{i}}(uco,F,W \ \& \ Rt)=-o_i^{t,\alpha_{i-1}}-o_i^{t,Z_{\alpha_{i-1},i}}$. \\

(c) \emph{$U_{i}$ griefs}: Since $U_{i-1}$ can claim a compensation of $Z_{\alpha_{i-1},i}$ by going on-chain and closing the channel. Payoff $u_{U_{i-1}}(uco,F,Gr)=-(o_{i-1}^{t_{i-1},\alpha_{i-1}}+o_{i-1}^{\tilde{t}_{i-1,i},remain(U_{i-1},U_i)}+M)+Z_{\alpha_{i-1},i}$.


$U_i$ incurs loss of $Z_{\alpha_{i-1},i}$ coins to compensate $U_{i-1}$. It fails to earn revenue due to non-utilization of $Z_{\alpha_{i-1},i}$ coins in channel $id_{i-1,i}$ for period $t_{i-1}$. The additional loses suffered are due to inability to claim $\alpha_{i-1}$ coins and using it within time $t_{i-1}$ and failure in utilizing the residual capacity $remain(U_{i},U_{i-1})$ for the next $T-(t_{i-1}+t_{contract\_initiate}-t_{i-1,i})$ units. Payoff of $U_i$ is $u_{U_{i}}(uco,F,Gr)=-o_{i}^{\tilde{t}_{i-1,i},remain(U_{i},U_{i-1})}-o_i^{t_{i-1},\alpha_{i-1}}-o_i^{t_{i-1},Z_{\alpha_{i-1},i}}-Z_{\alpha_{i-1},i}$.

\item[\textbf{B}.] If \textbf{N} had chosen an \emph{corrupt} node, then the payoffs are defined as follows:

(a) Instantaneous Response, i.e., $t \rightarrow 0$: Upon instant acceptance or rejection of payment, the payoffs are the same as that observed in $\Gamma_{HTLC}$.\\

(b) Delayed Response, i.e., $0<t\leq D-\delta$: If $U_{n}$ waits and then accepts the payment after $t$ units then payoff of $U_{n-1}$ is $u_{U_{n-1}}(co,F,W \ \& \ Ac)=f(v_{n-1})-o_{n-1}^{t,v_{n-1}}$. The loss observed is due to delay in claiming of $v_{n-1}$ coins as $U_n$ delays in resolving payment. $U_{n}$ keeps $v+Z_{v}\approx \alpha$ coins locked for mounting the attack and receives a bribe $L$. The value $\eta_n^{t,\alpha}$ is the same as defined in Eq.\ref{lock}. Upon accepting a self-payment of amount $v$, the corrupt node ends up paying a processing fee to $n-1$ intermediaries. Thus, the payoff of $U_{n}$ is $u_{U_{n}}(co,F,W \ \& \ Ac)=-\sum\limits_{i=1}^{n-1}f(v_i)+\eta_n^{t,\alpha}$.



If $U_{n}$ waits and then rejects the payment after $t$ units, then the payoff of $U_{n-1}$, $u_{U_{n-1}}(co,F,W \ \& \ Rt)=-o_{n-1}^{t,v_{n-1}}$ and $u_{U_{n}}(co,F,W \ \& \ Rt)=\eta_n^{t,\alpha}$. When $t \approx D-\delta$ where $\delta\rightarrow 0, \eta_n^{t,\alpha}$ attains the maximum value.\\

(c) \emph{$U_{n}$ griefs}: It gets an incentive $L$ from the attacker, but at the same time loses $Z_{v}$ in order to compensate the affected parties. $U_{n-1}$ loses channel and the expected revenue due to coins remaining locked but gets the compensation $Z_v$. Thus, the payoffs of $U_{n-1}$ and $U_n$ are $u_{U_{n-1}}(co,F,Gr)=-o_{n-1}^{D,v_{n-1}}-o_{n-1}^{\tilde{t}_{n-1,n},remain(U_{n-1},U_n)}-M+Z_v$ and $u_{U_{n}}(co,F,Gr)=L-C-o_n^{D,\alpha}-Z_v-o_{n-1}^{\tilde{t}_{n-1,n},remain(U_{n},U_{n-1})}=\eta_{n}^{D,\alpha}-Z_v-o_{n}^{\tilde{t}_{n-1,n},remain(U_{n},U_{n-1})}$ respectively.


\end{enumerate}

\subsubsection{Game Analysis}
\label{gp-analysisp}
If $U_{i-1}$ plays \emph{F}; an \emph{uncorrupt} $U_{i}$ chooses \emph{Ac} as its best response but a corrupt $U_{i}$ will choose \emph{Wait \& Reject} at $D-\delta$ as its best response since $u_{U_i}(co,W \ \& \ Rt \textrm{ at time }D-\delta)\geq u_{U_i}(co,F,s''), \forall s'' \in S_{U_i}$. The expected payoff of $U_{i-1}$ for selecting \emph{F}, denoted as $\mathbb{E}_{U_{i-1}}(F)$, is $\theta_i(-o_{n-1}^{D-\delta,v_{n-1}})+(1-\theta_i)f(\alpha_{i-1})$, and expected payoff for selecting \emph{NF}, denoted as $\mathbb{E}_{U_{i-1}}(NF)$, is 0. 
%

Since $\delta \rightarrow 0$, we consider $o^{D-\delta,v_{n-1}}_{n-1}\approx o^{D,v_{n-1}}_{n-1}$. If $\mathbb{E}_{U_{i-1}}(F)>\mathbb{E}_{U_i}(NF)$, then $U_{i-1}$ chooses \emph{F} else it aborts. We derive that $U_{i-1}$ chooses \emph{F} if $\theta_i<\frac{f(\alpha_{i-1})}{f(\alpha_{i-1})+o^{D,v_{n-1}}_{n-1}}$, else it chooses \emph{NF}; \emph{corrupt} $U_{i}$ chooses \emph{Wait \& Reject} at time $D-\delta$; \emph{uncorrupt} $U_{i}$ chooses \emph{Accept}; is a perfect Bayesian equilibrium.

\emph{Comparing $\theta_i$ for which $U_{i-1}$ chooses \emph{F} in $\Gamma_{HTLC}$ and $\Gamma_{HTLC-GP}$:} Since $f(\alpha_{i-1})+o^{D,v_{n-1}}_{n-1}<o^{D,\alpha_{n-1}}_{n-1}+f(\alpha_{i-1})+(1-q)(o_{n-1}^{\tilde{t}_{n-1,n},remain(U_{n-1},U_n)}+M)$, the cut-off of $\theta_i$ for which $U_{i-1}$ will choose to forward a payment is higher in $\Gamma_{HTLC-GP}$ than in $\Gamma_{HTLC}$ even if $q\rightarrow 0$. The corrupt player has to invest some amount as penalty and as a consequence, the payment amount reduces from $\alpha_{n-1}$ to $v_{n-1}$. Additionally, the corrupt player chooses to cancel the payment with $U_{n-1}$ off-chain just before elapse of locktime. This results in less risk compared to \emph{HTLC}. 


\subsection{Effectiveness of HTLC-GP}
\label{impact}
The analysis in Section \ref{gp-analysisp} shows that a rational corrupt node will cancel the payment off-chain just before the contract lock time elapses i.e., at time $D-\delta$, where $\delta\rightarrow 0$. The corrupt node avoids paying any penalty but still manages to mount the attack. We cannot protect uncorrupt parties from griefing attacks unless we do not account for the intermediate delay in resolving payments. Since Bitcoin scripting language is not Turing-complete, we cannot have a single off-chain contract where we can define penalty as a function of time. There is no way to execute a transaction like this: \emph{If t' time units have elapsed, pay amount p. If $t'+1$ time units have elapsed, pay amount $p+\delta$.} CheckSequenceVerify \cite{timelock} imposed on the first condition of elapse of $t'$ time unit makes the transaction eligible for broadcasting on-chain after elapse of time $t'+1$. This might lead to a race condition, and the victim might not receive proper compensation. Instead, it is desirable to construct $t'$ off-chain contracts, each accounting for a delay after every $t'$ interval. The timeout period of the $i^{th}$ contract is $i\frac{D}{t'}, i \in [1,t']$. However, multiple off-chain contracts for a single payment reduce the network throughput. Additionally, it is risky to have off-chain contracts with a shorter timeout period as it will lead to the abrupt closure of the payment channel and compromise the security of the protocol \cite{mizrahi2020congestion}. \emph{We conjecture that it is impossible to design an efficient protocol that will penalize the attacker and compensate the victims of a griefing attack with the current Bitcoin scripting system}.

Instead of focussing on the victims, we analyze the protocol from the attacker's point of view. The latter has an objective of maximizing the damage by locking as much network liquidity possible for the given budget $\mathcal{B}_{EX}$. The attacker will continue to invest in the network if the return on investment is good enough. If the return on investment diminishes, the attacker will refrain from mounting the attack and instead prefer to invest in another activity. In \emph{HTLC-GP}, the introduction of penalty led to locking extra coins, increasing the cost of the attack. The attacker will be able to corrupt fewer nodes compared to \emph{HTLC}. We define a metric, \emph{capacity locked in a path routing payment}, that indirectly determines the success rate of the attack \cite{bank}, \cite{lu2020general}. It is the summation of the coins locked in the off-chain contract instantiated on the channel forming the path. Ignoring the processing fee (negligible quantity), assuming all the payments executed are of value $\alpha$ and the bribe offered per instance is $L$, the attacker can corrupt $\frac{\mathcal{B}_{EX}}{L}$ nodes in the networks. We assume that for any node $U \in V$, $\mathbb{E}_{U}(\textrm{F})> \mathbb{E}_{U}(\textrm{NF})$. So each self-payment gets routed and reaches the payee.



\begin{claim}
\label{cl1}
\emph{\textcolor{black}{Given the total budget of the attack is $\mathcal{B}_{EX}$, incentive per attack being $L$, transaction value per payment being $\alpha$, HTLC timeout period is $D$, time taken to settle a transaction on-chain being $\Delta$, $n$ is the maximum allowed path length and a corrupt recipient rejects the payment at time $t'=D-\delta$, where $\delta \rightarrow 0$, the capacity locked upon using HTLC-GP is less than the capacity locked in HTLC, the loss percent being $\frac{\gamma n (\frac{D}{2}+\frac{\Delta (n-2)}{6})}{1+\gamma n (D+\frac{(n-1)\Delta}{2})}$}}
\end{claim}
\begin{claimproof}
In \emph{HTLC}, a given instance of attack locks $(n-1)\alpha$ coins in the path routing payment. The capacity locked in $\frac{\mathcal{B}_{EX}}{L}(n-1)\alpha$. 

In \emph{HTLC-GP}, $U_{n}$ executes self-payment of amount $v$. It cancels the contract at time $t'=D-\delta$. Capacity locked is $((n-1)v +\sum\limits_{i=1}^{n-1}Z_{v_{i-1},i})\frac{\mathcal{B}_{EX}}{L}$. In both the cases, we exclude the coins locked by the corrupt node while computing the unusable capacity and fee charged by the intermediate parties. Thus we substitute $Z_{v_{i-1},i}$ with $Z_{v,i}$. We measure the difference in capacity locked to judge the effectiveness.

\setlength\abovedisplayshortskip{0pt}
\begin{equation}
\begin{matrix}
    \frac{\mathcal{B}_{EX}}{L} (n-1)\alpha-\big(\frac{\mathcal{B}_{EX}}{L}(n-1)v + \frac{\mathcal{B}_{EX}}{L}\sum\limits_{i=1}^{n-1}Z_{v,i}\big)\\
    = \frac{\mathcal{B}_{EX}}{L} ((n-1)\alpha -(n-1)v -\sum\limits_{i=1}^{n-1}Z_{v,i})\\
        = \frac{\mathcal{B}_{EX}}{L} ((n-1)\alpha -(n-1)v - \gamma v \sum\limits_{i=0}^{n-1}\sum\limits_{j=0}^{i}t_j)\\
      \end{matrix}
\end{equation}
\setlength\belowdisplayshortskip{0pt}
Substituting $v=\frac{\alpha}{1+\gamma \sum\limits_{j=0}^{n-1} t_j}$,
\begin{equation}
\begin{matrix}
=\gamma \frac{\alpha}{1+\gamma (nD+\frac{n(n-1)\Delta}{2})} n(n-1)\frac{\mathcal{B}_{EX}}{L} (\frac{D}{2}+\frac{\Delta (n-2)}{6})
      \end{matrix}
\end{equation}

The loss percent is ratio of difference of capacity locked in \emph{HTLC} and \emph{HTLC-GP} and capacity locked in \emph{HTLC}.

\begin{equation}
\begin{matrix}
\frac{\gamma \frac{\alpha}{1+\gamma (nD+\frac{n(n-1)\Delta}{2})} n(n-1)\frac{\mathcal{B}_{EX}}{L} (\frac{D}{2}+\frac{\Delta (n-2)}{6}) }{    \frac{\mathcal{B}_{EX}}{L} (n-1)\alpha} \\\\
=\frac{\gamma n (\frac{D}{2}+\frac{\Delta (n-2)}{6})}{1+\gamma n (D+\frac{(n-1)\Delta}{2})}
\end{matrix}
\end{equation}

The loss percent of capacity locked by the attacker is $\frac{\gamma n (\frac{D}{2}+\frac{\Delta (n-2)}{6})}{1+\gamma n (D+\frac{(n-1)\Delta}{2})}$. The value is greater than 0 for $n\geq 2$. This proves that attacker ends up locking less capacity when HTLC-GP is used as a payment protocol.
\end{claimproof}

We observe that the loss percent $\frac{\gamma n (\frac{D}{2}+\frac{\Delta (n-2)}{6})}{1+\gamma n (D+\frac{(n-1)\Delta}{2})}$ is dependent on $\gamma$ \footnote{\textcolor{black}{It may not be always possible for a corrupt node to find a path of maximum allowed length for mounting the attack. In that case, the attacker might ask the corrupt node to get a the longest feasible path for routing payment. In our analysis, we consider the worst-case scenario where the corrupt node is able to route all its transaction via paths of length $n$.}}. If $\gamma$ is too low, the loss percent is not substantial and the attacker can still consider stalling the network. Hence the payment protocol \emph{HTLC-GP} is \emph{weakly effective} in disincentivizing the attacker. 

\begin{figure}[!ht]
     \centering
     \subfloat[Payoff of $U_{n-1}$, varying transaction value]{{\includegraphics[width=7cm]{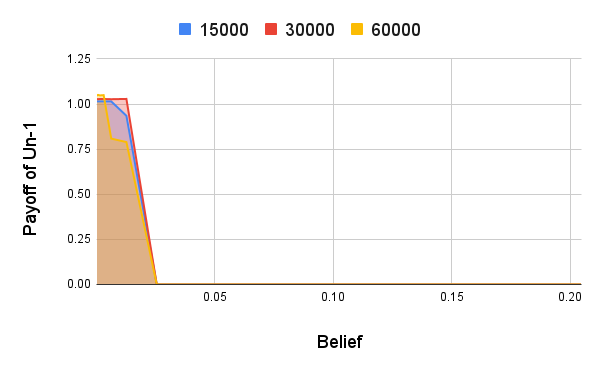} }}%
     \qquad
     \subfloat[Payoff of $U_{n}$, varying transaction value]{{\includegraphics[width=7cm]{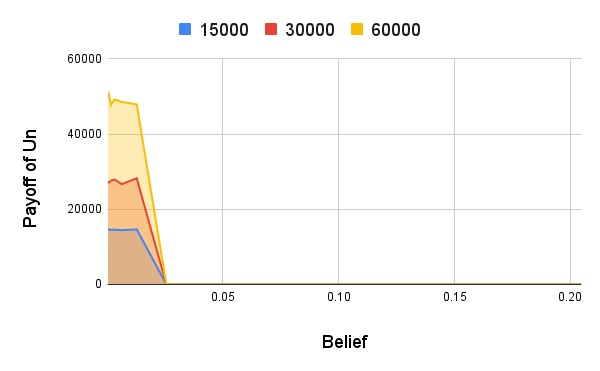} }}%
     
          \subfloat[Payoff of $U_{n-1}$, varying rate of arrival of transaction]{{\includegraphics[width=7cm]{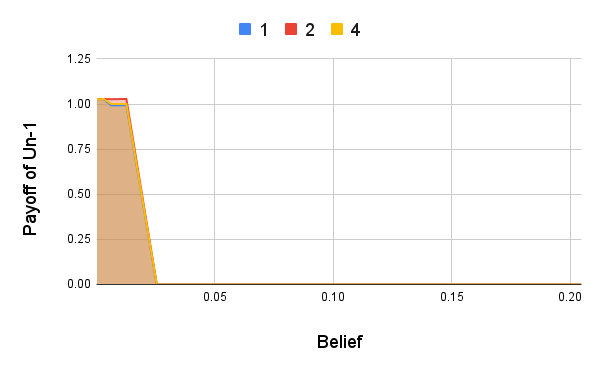} }}%
     \qquad
     \subfloat[Payoff of $U_{n}$,  varying rate of arrival of transaction]{{\includegraphics[width=7cm]{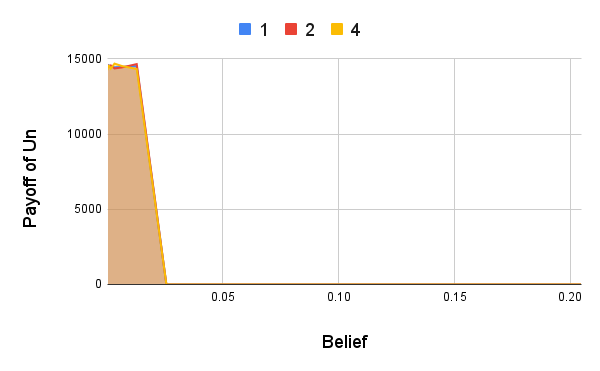} }}%
             \qquad
     \caption{Simulation of $\Gamma_{HTLC}$}%
     \label{fig:path}%
\end{figure}

 \begin{figure}[!ht]
     \centering
          \subfloat[Payoff of $U_{n-1}$, varying transaction value]{{\includegraphics[width=7cm]{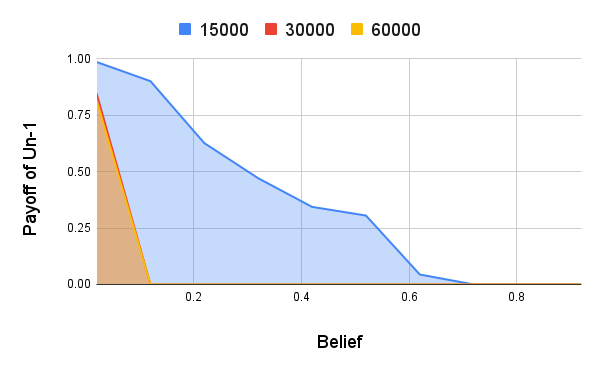} }}%
     \qquad
     \subfloat[Payoff of $U_{n}$, varying transaction value]{{\includegraphics[width=7cm]{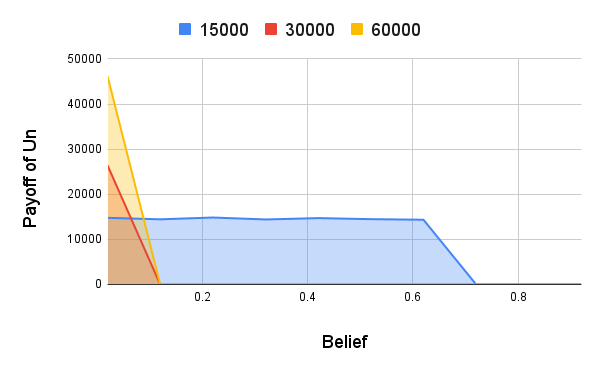} }}%
     
          \subfloat[Payoff of $U_{n-1}$, varying rate of arrival of transaction]{{\includegraphics[width=7cm]{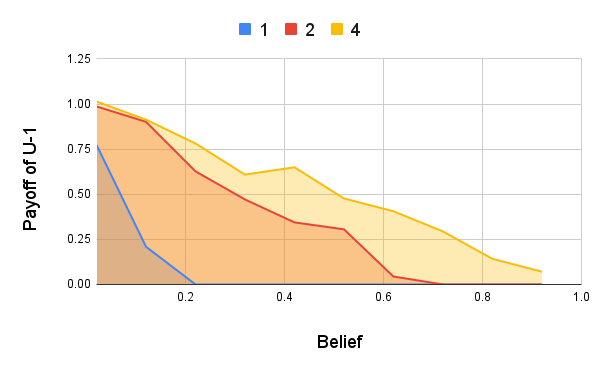} }}%
     \qquad
     \subfloat[Payoff of $U_{n}$,  varying rate of arrival of transaction]{{\includegraphics[width=7cm]{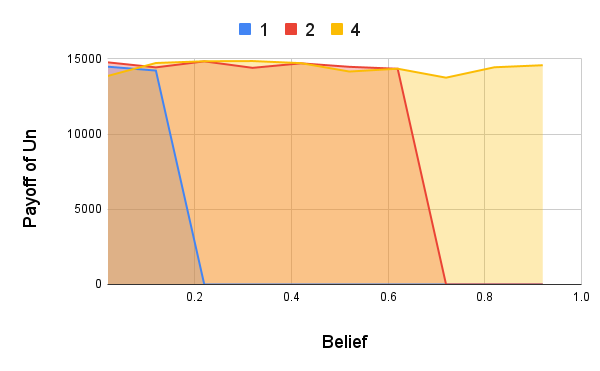} }}%
             \qquad

     \caption{Simulation of $\Gamma_{HTLC-GP}$}%
     \label{fig:path1}%
\end{figure}

\section{Simulating the game model for griefing attack in HTLC and HTLC-GP}
\subsection{Dataset and Parameters}
\emph{Simulation of sequential Bayesian games}: The first part analyzes the payoff of each party involved in the games $\Gamma_{HTLC}$ and $\Gamma_{HTLC-GP}$. We simulate the games $\Gamma_{HTLC}$ and, $\Gamma_{HTLC-GP}$ respectively, and estimate the payoff of $U_{n-1}$ and $U_{n}$. We consider a \emph{Poisson distribution} for the arrival of transaction in a given channel \cite{kawase2017transaction}. The rate of arrival of the transaction is varied between 1 and 4 for the next $10\ blocks$. The path length is set 20 and $D=100$. The transaction amount is varied between $15000$ satoshis to $60000$ satoshis. The mining fee for closing a channel is $0.00000154 \ BTC$\footnote{We have considered the data for mining fee observed for one particular channel closure \url{https://blockstream.info/tx/c0471c9ff72a883aa45058029049ffa12b92d7379f44447bc1df52382c725c01}, the mining fee can vary as observed for various closed channels in \url{https://1ml.com/channel?order=closedchannels}}. $q$ is set to $0.7$. If the coins remaining unutilized are $C$, the party tries to estimate the fee earned in the future had it utilized the coins. We set $per\_tx\_val$ to 1000 satoshis, thus a party will earn by processing $\frac{C}{1000}$ transactions.
\subsection{Observations} 
We discuss our observation in this section:
\begin{itemize}
\item  $\Gamma_{HTLC}$: For transaction value ranging $15000 - 60000$ (in satoshis) and rate of arrival of transaction fixed to 10 within 10 blocks , the plots in Fig. \ref{fig:path}(a) and \ref{fig:path}(b) shows the expected payoff of $U_{n-1}$ and expected payoff of $U_{n}$ varying with the belief $\theta$. $U_{n-1}$'s payoff decreases with increase in $\theta$. Payoff of $U_{n}$ remains more or less constant for a fixed transaction amount, but increases with increase in the transaction amount. For $\theta\geq 0.025$, $U_{n-1}$ acts cautious and chooses \emph{not forward}, as forwarding will lead to negative payoff. Both $U_{n-1}$'s and $U_{n}$'s payoff drops to $0$ from this point onwards.

In Fig. \ref{fig:path}(c) and \ref{fig:path}(d), the rate of arrival of transaction is varied between 1 and 4 within a period of 10 blocks and transaction amount is 15000 satoshi. We observe that for $\theta< 0.025$, payoff of $U_{n-1}$ and $U_n$ remains positive and invariant.

\item $\Gamma_{HTLC-GP}$: The plots in Fig. \ref{fig:path1}(a) and \ref{fig:path1}(b) shows that for $\theta\geq 0.1$, $U_{n-1}$ acts cautious and chooses \emph{not forward} for transaction varying between 30000 satoshi and 60000 satoshis. For transaction amount 15000, expected payoff of $U_{n-1}$ and $U_n$ remains positive till $\theta<0.7$.

In Fig. \ref{fig:path1}(c) and \ref{fig:path1}(d), given the rate of arrival of transaction is 1,  $U_{n-1}$ chooses to \emph{forward} till $\theta\leq 0.2$. When the rate of arrival of transaction is 2, $U_{n-1}$ chooses to \emph{forward} till $\theta\leq 0.7$ and when rate of arrival is 4, $\theta\leq 0.9$.

\end{itemize}
\subsection{Discussion of Results}
\emph{Expected Payoff in $\Gamma_{HTLC}$ and $\Gamma_{HTLC-GP}$}: 
\begin{itemize}[leftmargin=*]

\item \emph{Transaction amount is varied}: We see that expected payoff of $U_{n-1}$ in $\Gamma_{HTLC-GP}$ remains positive for a higher value of $\theta$ compared $\Gamma_{HTLC}$. The reason being that in the first game, $U_n$ can choose not to respond, forcing $U_{n-1}$ to go on-chain and close the channel. Since the mining fee for closing the channel is quite high, the stakes are higher. Thus $U_{n-1}$ tends to stop forwarding payment for $\theta$ as low as $0.025$. In the second game, $U_n$ will always resolve the payment just before the lock time elapses to avoid paying penalty. This prevents abrupt closure of the channel. It is observed that the cutoff value of $\theta$ is higher for transaction amount $150000$ satoshis. This is because the capacity locked is higher when the transaction amount increases, hence the risk is higher.

\item \emph{Rate of the arrival of the transaction is varied}: In $\Gamma_{HTLC}$, varying the rate of arrival of the transaction has no impact on the payoff of both $U_{n-1}$ and $U_{n}$ because mining fee of channel closure dominates the result. In $\Gamma_{HTLC-GP}$, the value of $\theta$ increases with an increase in the rate of arrival of transaction. The reason behind this is that the distribution is positively skewed for a lower arrival rate. As the rate increases, the Poisson distribution becomes more symmetric and less peaked.

\end{itemize}
 
\section{Guaranteed minimum compensation}
\label{guarantee}
If the incidence of griefing attack increases in the network, $\gamma$ can be increased. However, the disadvantage of increasing the rate of griefing penalty means uncorrupt nodes have to put a higher amount at stake for routing small-valued payments. The success rate of payments decreases due to a lack of liquidity in the channels. Our objective is to increase the cost of the attack without forcing uncorrupt parties to lock high penalties. \textcolor{black}{Since corrupt parties are asked to route self-payment via the longest available path, cost of the attack increases if the maximum path length allowed for routing payments decreases \cite{mizrahi2020congestion}. This would lead to locking of less coins in the network by the corrupt parties. However, an abrupt reduction in the maximum allowed path length may lead to higher failure in executing transactions. Thus we must design a mechanism by which one can adjust the maximum allowed path length based on the rate of incidence of griefing attacks in the network.}

The major source of earning for a node is the processing fee by routing transactions. If there is a griefing attack, then the affected parties fail to earn due to locked collateral. We introduce a new parameter $\zeta$, termed as \emph{Guaranteed Minimum Compensation}. Based on the data provided \cite{beres2019cryptoeconomic}, the fee earned by each node on a single day is quite low compared to the amount routed. Thus, we set $\zeta$ in the range $[0,1)$  to avoid over-compensation.

\subsubsection*{Adjusting the parameters based on $\zeta$}

\textcolor{black}{Let the maximum allowed path length in the new model be denoted as $\tilde{n}^{\zeta,k}$. If payment forwarded by $U_{i-1}$ is $\alpha$, $U_i$ must lock a minimum cumulative penalty $i \zeta \alpha, i \in [1, \kappa]$ where $\kappa \leq \tilde{n}^{\zeta,k} $. Each party $U_j, j \in [0,i)$ is entitled to receive a compensation $\zeta\alpha$. For a given rate of penalty, let the maximum cumulative penalty an uncorrupt recipient has to lock when payment is routed via path of length $\tilde{n}^{\zeta,k}$ be $Z_{\alpha,max}=k\alpha$, where $k\in \mathbb{R}^{+}$.}

If the probability of remaining live is $h$ and $(1-h)$ is the probability of suffering crash fault, then the payee would be willing to lock a maximum cumulative griefing-penalty $Z_{\alpha,max}=k\alpha$, such that the expected profit is greater than 0. Given that it gains $\alpha$ when it remains live, and loses $k\alpha+o_n^{D,k\alpha}+o_n^{D,\alpha}+o_n^{\tilde{t}_{n-1,n},remain(U_n,U_{n-1})}$ if there is a crash fault, the expected profit if the payee participates in the game is $h\alpha-(1-h)\big(k\alpha+o_n^{D,k\alpha}+o_n^{D,\alpha}+o_n^{\tilde{t}_{n-1,n},remain(U_n,U_{n-1})}\big)$ and the expected profit if it does not participate is 0. If a payee participates in the game, then the following relation must hold:
\begin{equation}
\begin{matrix}
    h\alpha-(1-h)\big(k\alpha+o_n^{D,k\alpha}+o_n^{D,\alpha}\\ +o_n^{\tilde{t}_{n-1,n},remain(U_n,U_{n-1})}\big)\geq 0\\
\end{matrix}
\end{equation}
From the above equation, the payee calculates the maximum value of $k$ for which it would be safe to participate in the game. Given the conditions, we discuss how to calculate the maximum allowed path length for a payment and its corresponding rate of griefing-penalty.

\begin{itemize}
    \item[a.] If all the nodes from $U_0$ to $U_{\tilde{n}^{\zeta,k}-1}$ charge a minimum compensation of $\zeta \alpha$, then the minimum value of $Z_{\alpha,max}$ is $\tilde{n}^{\zeta,k}\zeta \alpha$. We use this relation to calculate the maximum allowed path length $\tilde{n}^{\zeta,k}$ for routing payment.
\begin{proposition}
\label{pp1}
\emph{Given the maximum cumulative griefing-penalty for a payment $\alpha$ is $k\alpha$, and the guaranteed minimum compensation $\zeta$, the maximum allowed path length $\tilde{n}^{\zeta,k}$ is $\frac{k}{\zeta}$.}
\end{proposition}
\begin{claimproof}
In \emph{HTLC-GP}, $Z_{\alpha,max}=k\alpha$ when payment is routed via the maximum allowed path length $\tilde{n}^{\zeta,k}$. Upon introduction of \emph{guaranteed minimum compensation}, any node routing a payment $\alpha$ is entitled to a minimum compensation $\zeta \alpha$ upon being affected by the griefing attack. If the compensation falls below this amount, the node will refuse to forward the contract. Thus, the recipient must bear a cumulative penalty of at least $\zeta \tilde{n}^{\zeta,k}\alpha$ for a path of length $\tilde{n}^{\zeta,k}$. Thus, the following criteria must hold:
\setlength\abovedisplayshortskip{0pt}
\begin{equation}
\label{b1}
\begin{matrix}
\tilde{n}^{\zeta,k} \zeta \alpha  \leq Z_{\alpha,max} \\\\
or, \tilde{n}^{\zeta,k} \zeta \alpha  \leq k\alpha \\\\
or, \tilde{n}^{\zeta,k} \leq \frac{k}{\zeta}
 \end{matrix}
 \end{equation}
 Thus the maximum path length in $\textrm{HTLC-GP}^{\zeta}$ is $\frac{k}{\zeta}$.
\end{claimproof}
From the expression, we observe that $\tilde{n}^{\zeta,k}$ is inversely proportional to $\zeta$. 


    \item[b.] Once the maximum path length is adjusted, the rate of griefing-penalty $\gamma$ ceases to be an independent variable. We call this new rate of griefing-penalty $\gamma^{\zeta,k}$ and provide an expression for calculating the same.

\begin{proposition}
\label{pp2}
\emph{Given the guaranteed minimum compensation $\zeta$, ratio of maximum cumulative griefing penalty and the transaction amount is $k$, and \emph{HTLC-GP} timeout period $D$, the rate of griefing-penalty $\gamma^{\zeta,k}$ is $\frac{2\zeta^2}{2\zeta D + \Delta(k-\zeta)}$.}

\end{proposition}
\begin{claimproof} 
For ease of analysis, we ignore the processing fee charged by each intermediate party. The cumulative penalty locked by the recipient is summation of the compensation of each intermediary and the source node forwarding the conditional payment. Given that $Z_{\alpha,max}=k \alpha$ and maximum path length for routing is $\tilde{n}^{\zeta,k}$, we estimate $\gamma^{\zeta,k}$:
\setlength\abovedisplayshortskip{0pt}

\begin{equation}
\label{zeta2}
\begin{matrix}
 k\alpha=\gamma^{\zeta,k}\sum_{i=1}^{\tilde{n}^{\zeta,k}} (D+({\tilde{n}^{\zeta,k}}-i)\Delta)\alpha\\
or,  \qquad \gamma^{\zeta,k} = 
  \frac{k}{ \Delta {\tilde{n}^{\zeta,k}} (\frac{D}{\Delta}+\frac{{\tilde{n}^{\zeta,k}}-1}{2})}
\end{matrix}
\end{equation}
\setlength\belowdisplayshortskip{0pt}

Replacing $\tilde{n}^{\zeta,k}$ by the expression given in Proposition \ref{pp1}, we have $\gamma^{\zeta,k}=\frac{2\zeta^2}{2\zeta D + \Delta(k-\zeta)}$.
\end{claimproof}

\end{itemize}

\section{Modifying HTLC-GP to HTLC-GP$^{\zeta}$}
\label{modify}

Given a payment request $(S,R,\alpha)$, $R$ decides on the maximum cumulative penalty $k\alpha$ for a given $\zeta$. \textcolor{black}{$S$ find a path of length $\kappa$ where $\kappa \leq \tilde{n}^{\zeta,\kappa}=\frac{k}{\zeta}$ for routing the payment where $U_0=S$ and $U_{\kappa}=R$. $t_{\kappa-1}$ is set to $D$ and the rate of griefing-penalty $\gamma^{\zeta,k}$ is calculated accordingly. The total amount that the payer needs to transfer is $\tilde{\alpha}=\alpha+\sum\limits_{j=1}^{\kappa-1}f(\alpha_j)$. We denote each $\alpha_i=\tilde{\alpha}- \sum\limits_{k=1}^{i}f(\alpha_k), i \in [0,\kappa-1],  \alpha_0=\tilde{\alpha}$. Each node $U_i$ samples a pair of secret key and public key $(sk_i,pk_i)$, the public key of each node is used to encrypt the information of establishing contract with the neighboring node. }

\subsection{Protocol Description}
The phases of $\textrm{HTLC-GP}^{\zeta}$ is similar to \emph{HTLC-GP} \cite{agrief}.\\
\noindent \texttt{(a)Pre-processing Phase}: If the exact path length $\kappa$ is used for routing payment, $U_1$ locks a penalty $\gamma^{\zeta,k} \alpha_0 t_0$ with $U_0$ and the former can easily figure out the identity of the sender. To prevent violation of privacy, $U_0$ randomizes the exact path length using a random function $\phi$, and shares $\phi(\kappa)$ with $U_{\kappa}$. The latter calculates the cumulative penalty $\gamma^{\zeta,k} \phi(\kappa) \alpha D$ used for establishing the \emph{cancellation contract}. A routing attempt cost $\psi$ is added such that $\gamma^{\zeta,k} \phi(\kappa) \alpha D \approx  \gamma^{\zeta,k} ((\psi + \alpha_{0})t_{0}+\Sigma_{j=1}^{\kappa-1}\alpha_{j}t_{j})$. This acts like a blinding factor and $U_1$ cannot infer the identity of the sender from the penalty it locks in the off-chain contract formed with $U_0$. The following steps of the protocols are executed:\\
(i) $U_{\kappa}$ samples two random numbers $x$ and $r$ where $x\neq r$. It  constructs the payment hash $H=\mathcal{H}(x)$ and the cancellation hash $Y=\mathcal{H}(r)$. \\
(ii) The payee shares both the hashes $H$ and $Y$ with the $U_0$. The cumulative griefing-penalty to be locked by $U_1$ is cgp$_{0}=\gamma^{\zeta,k} (\psi+\tilde{\alpha})t_0$. The cumulative griefing-penalty to be locked by any other node $U_{i+1}, i\in [1,\kappa-1]$ is cgp$_{i}= \gamma^{\zeta,k}.(\Sigma_{j=1}^{i} (\alpha_j t_j)+(\alpha_0+\psi)t_0)$.\\
(iii) The payer uses standard onion routing \cite{goldschlag1999onion} for propagating the information needed by each node $U_i, i \in [1,\kappa]$, across the path $P$. $U_0$ sends $M_0=E(\ldots E(E(E(\phi,Z_{\kappa},pk_{\kappa}),Z_{\kappa-1},pk_{\kappa-1}), Z_{\kappa-2},pk_{\kappa-2}),Z_{1},pk_{1})$ to $U_1$, where $Z_i=(H,Y,\alpha_i, t_{i-1},\textrm{cgp}_{i-1},U_{i+1}), i \in [1,\kappa-1]$ and $Z_{\kappa}=(H,Y,\alpha_{\kappa-1},t_{\kappa-1}, \textrm{cgp}_{\kappa-1},null)$. Here $M_{i-1}=E(M_{i},Z_{i},pk_{i})$ is the encryption of the message $M_{i}$ and $Z_{i}$ using public key $pk_{i}$, $M_{\kappa}=\phi$. \\
(iv) $U_1$ decrypts $M_0$, gets $Z_1$ and $M_1$. $M_1=E(\ldots E(E(E(\phi,Z_{\kappa},pk_{\kappa}),Z_{\kappa-1},pk_{\kappa-1}), Z_{\kappa-2},pk_{n-2}),\ldots, Z_{2},\\pk_{2})$ is forwarded to the next destination $U_2$. This continues till party $U_{\kappa}$ gets $E(\phi,Z_{\kappa},pk_{\kappa})$.\\  

%

\noindent \texttt{(b) Two-Round Locking Phase}: It involves the following two rounds:
\begin{itemize}
\item \textit{Establishing Cancellation Contract}: $U_{\kappa}$ initiates this round and each player $U_i, i\in [1,\kappa]$ locks their respective cumulative griefing-penalty $cgp_{i-1}$. 
\begin{itemize}
\item[(i)] $U_{\kappa}$ decrypts and gets $Z_{\kappa}$. It checks $\gamma^{\zeta,k} \phi(\kappa) \alpha D \stackrel{?}{=}cgp_{\kappa-1}$ and $\alpha_{\kappa-1}\stackrel{?}{=}\alpha$. If this holds, the payer sends a contract formation request to $U_{\kappa-1}$. The latter knows that it has to lock $\alpha_{\kappa-1}$ with $U_{\kappa}$ in the second round, so it checks that given the belief $\theta_{\kappa}$ regarding $U_{\kappa}$'s type, the expected payoff on forwarding the contract in second round is greater than 0. If so, it accepts the terms of the off-chain contract from $U_{\kappa}$, with the latter locking $cgp_{\kappa-1}$ coins. 
\item[(ii)] For any other party $U_{i}, i \in [1,\kappa-1]$, it first checks $cgp_{i}-\gamma^{\zeta,k} \alpha_{i} t_{i} \stackrel{?}{=} cgp_{i-1}$. This ensures that there is sufficient coins to be locked as penalty in the contract to be formed with $U_{i-1}$. Next, it checks $\gamma^{\zeta,k} \alpha_i t_i \geq \zeta \alpha_i$. This check ensures that $U_i$ is guaranteed a minimum compensation upon being affected by griefing attack. If both the condition satifies, $U_i$ sends a request to form off-chain contract with $U_{i-1}$. If the latter accepts based on the belief $\theta_i$ regarding $U_i's$ nature, $U_i$ locks $cgp_{i-1}$. 
\item[(iii)] The off-chain contract for locking penalty in layman terms: \emph{`$U_{i+1}$ can withdraw the amount $cgp_i=\gamma^{\zeta,k}.(\Sigma_{j=1}^{i} (\alpha_j t_j)+(\alpha_0+\psi)t_0)$ from the contract contingent to the release of either $x: H=\mathcal{H}(x)$ or $\ r: Y=\mathcal{H}(r)$ within time $t_i$. If the locktime elapses and $U_{i+1}$ does not respond, $U_{i}$ claims $cgp_i$ after the locktime elapses.'}.  
\end{itemize}

The pseudocode of the first round of Locking Phase for $U_{\kappa}$, any intermediate party $U_i, i \in [1,\kappa-1]$ and payer $U_0$ is stated in Procedure \ref{algo:lock}, Procedure \ref{algo:lock1} and Procedure \ref{algo:lock2} respectively.


\begin{proc}[!ht]
    \SetKwInOut{Input}{Input}
    \SetKwInOut{Output}{Output}

    \caption{Establishing Cancellation Contract: First Round of Locking Phase for $U_{\kappa}$ }
        \label{algo:lock}
        
\textbf{Input}: $(Z_{\kappa},\phi(\kappa),\gamma^{\zeta,k}, \alpha)$  \\
$U_{\kappa}$ parses $Z_{\kappa}$ and gets $H', Y',\alpha',t',\textrm{cgp}_{\kappa-1}$.\\
\If{$t'\geq t_{now}+\Delta$ and $\alpha'\stackrel{?}{=}\alpha$ and $k \alpha \stackrel{?}{=} \textrm{cgp}_{\kappa-1}$ and $H'\stackrel{?}{=}H$ and $Y'\stackrel{?}{=}Y$  and $remain(U_{\kappa},U_{\kappa-1})\geq \textrm{cgp}_{\kappa-1}$}
{

    Send $\textrm{Cancel\_Contract\_Request}(H,Y,t',\textrm{cgp}_{\kappa-1},\gamma^{\zeta,k})$ to $U_{\kappa-1}$\\
     \If{acknowledgement received from $U_{\kappa-1}$}
{      
  $remain(U_{\kappa},U_{\kappa-1})=remain(U_{\kappa},U_{\kappa-1})-\textrm{cgp}_{\kappa-1}$\\
  establish $Cancel\_Contract(H,Y,t',\textrm{cgp}_{\kappa-1})$ with $U_{\kappa-1}$\\
     Record $t_{\kappa}^{form}=current\_clock\_time$\\
     }
     \Else
     {
       abort
     }

 }
 \Else
{

 abort.
}
 
 \end{proc}

\begin{proc}[!ht]
    \SetKwInOut{Input}{Input}
    \SetKwInOut{Output}{Output}

    \caption{Establishing Cancellation Contract: First Round of Locking Phase for $U_i, i\in [1,\kappa-1]$ }
        \label{algo:lock1}
        
\textbf{Input}: $(H',Y',t',\textrm{cgp}_{i},\gamma^{\zeta,k})$  \\
$U_i$ parses $Z_{i}$ and gets $H, Y,\alpha_{i},t_{i-1},\textrm{cgp}_{i-1}$.\\

\If{$\theta_{i+1} < \frac{f(\alpha_i)}{f(\alpha_i)+o^{t_i,\alpha_i}_{i}}$ and $H'\stackrel{?}{=}H$ and $Y\stackrel{?}{=}Y'$ and $t'+\Delta\stackrel{?}{\leq} t_{i-1}$ and $\textrm{cgp}_{i}-\gamma^{\zeta,k} \alpha_i t'\stackrel{?}{=}\textrm{cgp}_{i-1}$ and $\gamma^{\zeta,k} \alpha_i t'\geq \zeta \alpha_i$ and $remain(U_i,U_{i+1})\geq \alpha_i$ and $remain(U_{i},U_{i-1})\geq \textrm{cgp}_{i-1}$ and (current\_time not close to contract expiration time)}
{

  Sends acknowledgment to $U_{i+1}$ and waits for the off-chain contract to be established\\
  
       Send $\textrm{Cancel\_Contract\_Request}(H,Y,t_{i-1},\textrm{cgp}_{i-1},\gamma^{\zeta,k})$ to $U_{i-1}$\\

     \If{acknowledgement received from $U_{i-1}$}
{      
  $remain(U_{i},U_{i-1})=remain(U_i,U_{i-1})-\textrm{cgp}_{i-1}$\\
  establish $Cancel\_Contract(H,Y,t_{i-1},\textrm{cgp}_{i-1})$ with $U_{i-1}$\\
     }
     \Else
     {
     
       abort
     }

}
\Else
{

 abort.
}
    
\end{proc}

\begin{proc}[!ht]
    \SetKwInOut{Input}{Input}
    \SetKwInOut{Output}{Output}

    \caption{Establishing Cancellation Contract: First Round of Locking Phase for $U_0$ }
        \label{algo:lock2}
        
\textbf{Input}: $(H',Y',t',\textrm{cgp}',\gamma^{\zeta,k})$  \\

\If{$\theta_1 < \frac{f(\alpha_0)}{f(\alpha_0)+o^{t_0,\alpha_0}_{0}}$ and $t'\stackrel{?}{=}t_0$ and $\textrm{cgp}'\stackrel{?}{=}\textrm{cgp}_0\geq \zeta \alpha_0$ and $H'\stackrel{?}{=}H$ and $Y'\stackrel{?}{=}Y$  and $remain(U_{0},U_{1})\geq \alpha_0$}
{

  Sends acknowledgment to $U_{1}$\\
  Confirm formation of penalty contract with $U_1$\\
  Initiate the second round, the establishment of payment contract\\

 }
 \Else
{

 abort.
}
 
 \end{proc}

\item \textit{Establishing Payment Contract}: $U_0$ initiates the next rounded provided it has received the cancellation contract and $cgp_0\geq \zeta \alpha_0$. The conditional payment is forwarded till it reaches the payer $U_{\kappa}$. This proceeds as normal \emph{HTLC}. 
\begin{itemize}
\item[(i)] A node $U_i, i \in [0,\kappa-1]$ forms the off-chain payment contract with $U_{i+1}$, locking $\alpha_i$ coins, if and only if $U_i$ had accepted the formation of cancellation contract with $U_{i+1}$ in the first round.
\item[(ii)]  The off-chain contract for payment in layman terms: \emph{`$U_{i+1}$ can claim $\alpha_i$ coins contingent to the release of $x:H=\mathcal{H}(x)$ within time $t_i$. If $U_{i+1}$ does not respond, $U_{i}$ unlocks $\alpha_i$ coins from the contract either by releasing preimage $r:Y=\mathcal{H}(r)$ or after the locktime elapses.'}
\end{itemize}
\end{itemize}

The pseudocode of the second round of Locking Phase for a party $U_i, i \in [0,\kappa-1]$ is stated in Procedure \ref{algo:lock4}.

 \begin{proc}[H]
    \SetKwInOut{Input}{Input}
    \SetKwInOut{Output}{Output}

    \caption{Establishing Payment Contract: Second Round of Locking Phase for $U_i, i \in [0,\kappa-1]$ }
        \label{algo:lock4}
        
\textbf{Input}: $(H,Y,\alpha_i,t_i)$  \\

\If{ (cancellation contract locking penalty has been formed between $U_i$ and $U_{i+1}$ in the first round) and $t_{i-1}\geq t_i +\Delta$ and $\alpha_i\stackrel{?}{=}\alpha_{i-1}+fee(U_i)$ and ($U_{i+1}$ has agreed to form the contract) and (current\_time not close to contract expiration time)}
{
  $remain(U_{i},U_{i+1})=remain(U_i,U_{i+1})-\alpha_i$\\
  establish $Payment\_Contract(H,Y,t_i,\alpha_i)$ with $U_{i+1}$\\
 }
 \Else
 {
 abort
 }
 \end{proc}

\noindent \texttt{(c) Release Phase}: $U_{\kappa}$ waits for a very short duration, say $\mu$, to receive the payment contract from $U_{\kappa-1}$. If the payment contract has been forwarded by $U_{\kappa-1}$ within $\mu$ units of time and it is correct, then $U_{\kappa}$ releases the preimage $x$ for payment hash $H$ and claims the coins from $U_{\kappa-1}$. If the latter has delayed beyond $\mu$, or the payment contract forwarded by $U_{\kappa-1}$ is invalid, $U_{\kappa}$ releases the cancellation preimage $r$. In case of dispute, the payer goes on-chain and releases one of the preimages for settling the contract. The rest of the parties $U_{i+1}, i \in [0,\kappa-2]$ either claim the coins or cancel the payment based on the preimage released. If $U_{i+1}$ griefs and refuses to release preimage to $U_i$, the former has to pay the cumulative griefing-penalty $cgp_{i}$ for affecting the nodes $U_k, 0\leq k \leq i$, so that all the nodes obtain their due compensation. We discuss the Release Phase of the protocol for node $U_{\kappa}$ and any intermediary $U_i,  i \in [1,\kappa-1]$ in Procedure \ref{algo:rel11} and Procedure \ref{algo:rel12}.

 \begin{proc}[H]
    \SetKwInOut{Input}{Input}
    \SetKwInOut{Output}{Output}

    \caption{Release\_Phase for $U_{\kappa}$}
        \label{algo:rel11}
        
\textbf{Input}: Message $M$, time bound $\mu$  \\

\If{$M\stackrel{?}{=}Payment\_Contract(H,Y,\alpha',t')$ and $current\_clock\_time-t_{\kappa}^{form}\leq \mu$}
{
Parse $M$ and retrieve $(H,Y,\alpha',t')$\\
   \If{ $t'\geq t_{now}+ \Delta$ and $\alpha'=\alpha$}
   {
      $z=x$\\
      
    }
    \Else
    {
      $z=r$\\
       
     }
   }
   \Else{
     $z=r$\\
   }
      Release $z$ to $U_{\kappa-1}$\\
      \If{$current\_time < t_{\kappa-1}$}
      {
         
         \If{$U_{\kappa}$ and $U_{\kappa-1}$ mutually agree to terminate \emph{Payment Contract and Cancellation Contract}}
         {
         
         \If{z=x}
         {
        $remain(U_{\kappa},U_{\kappa-1})=remain(U_{\kappa},U_{\kappa-1})+\alpha+\textrm{cgp}_{\kappa-1}$\\         
        }
        \Else
        {
          $remain(U_{\kappa},U_{\kappa-1})=remain(U_{\kappa},U_{\kappa-1})+\textrm{cgp}_{\kappa-1}$\\         
          $remain(U_{\kappa-1},U_{\kappa})=remain(U_{\kappa-1},U_{\kappa})+\alpha$\\         
        }
           
           }
           \Else
           {
  $U_{\kappa}$ goes on-chain for settlement by releasing preimage $z$.           

         }
         
      }
      \Else
      {
      $U_{\kappa-1}$ goes on-chain for settlement, claims $(\alpha+\textrm{cgp}_{\kappa-1})$. \\
      $z=null$\\
      }
                    Call Release\_Phase($U_{\kappa-1},z$)\\

   \end{proc}
\begin{proc}
    \SetKwInOut{Input}{Input}
    \SetKwInOut{Output}{Output}

    \caption{Release\_Phase for $U_i, i \in [1,\kappa-1]$}
        \label{algo:rel12}
        
\textbf{Input}: $z$  \\

      Release $z$ to $U_{i-1}$\\
      \If{$z\neq null$ and $current\_time < t_{i-1}$}
      {
         
         \If{$U_i$ and $U_{i-1}$ mutually agree to terminate \emph{Payment Contract and Cancellation Contract}}
         {
        \If{z=x}
         {
        $remain(U_{i},U_{i-1})=remain(U_{i},U_{i-1})+\alpha_{i-1}+\textrm{cgp}_{i-1}$\\         
        }
        \Else
        {
          $remain(U_i,U_{i-1})=remain(U_i,U_{i-1})+\textrm{cgp}_{i-1}$\\         
          $remain(U_{i-1},U_i)=remain(U_{i-1},U_i)+\alpha_{i-1}$\\         
        }

           }
           \Else
           {
  $U_i$ goes on-chain for settlement by releasing preimage $z$.\\

         }
         }
         
\Else
{
      $U_{i-1}$ goes on-chain for settlement after elapse of locktime $t_{i-1}$, claims $(\alpha_{i-1}+\textrm{cgp}_{i-1})$. \\
}

                    Call Release\_Phase($U_{i-1},z$)\\

   \end{proc}

\subsection{Effectiveness of HTLC-GP$^{\zeta}$}
A corrupt node can still mount the attack by canceling the payment just before the off-chain contract's lock time elapses. However, we intend to study the impact of the reduced maximum allowed path length $\textcolor{black}{\tilde{n}^{\zeta,k}}$ on the effective \emph{capacity locked} by the attacker in the network. We assume that for any node $U \in V$, $\mathbb{E}_{U}(\textrm{F})> \mathbb{E}_{U}(\textrm{NF})$ and thus each self-payment gets routed and reaches the payee.

\begin{claim}
\label{cl2}
\emph{Given the total budget of the attack is $\mathcal{B}_{EX}$, incentive per attack being $L$,  transaction value per payment being $\alpha$, HTLC timeout period is $D$, time taken to settle a transaction on-chain being $\Delta$, $n$ is the maximum allowed path length for HTLC, \textcolor{black}{$\tilde{n}^{\zeta,k}$ is the maximum allowed path length for $\textrm{HTLC-GP}^{\zeta}$ for a given pair of $\zeta$ and $k$}, and a corrupt recipient rejects the payment at time $t'=D-\mu$, where $\mu \rightarrow 0$, the capacity locked in $\textrm{HTLC-GP}^{\zeta}$ is less than the capacity locked in HTLC, the loss percent being $\frac{n-\tilde{n}^{\zeta,k}}{(n-1)\Big(1+\gamma^{\zeta,k} n D+ \gamma^{\zeta,k} n \Delta \frac{n-1}{2}\Big)}+\frac{\gamma^{\zeta,k} \tilde{n}^{\zeta,k} ((n-1)(D+\frac{(\tilde{n}^{\zeta,k}-1)\Delta}{2})-\frac{\tilde{n}^{\zeta,k}-1}{2}(D+\frac{(2\tilde{n}^{\zeta,k})\Delta}{3}))}{(n-1)\Big(1+\gamma^{\zeta,k} n D+ \gamma^{\zeta,k} n \Delta \frac{n-1}{2}\Big)}$}
\end{claim}
\begin{claimproof}
In $\textrm{HTLC-GP}^{\zeta}$, the capacity locked is $((\tilde{n}^{\zeta,k}-1)v +\sum\limits_{i=1}^{\tilde{n}^{\zeta,k}-1}Z_{v,i})\frac{\mathcal{B}_{EX}}{L}$. In both the cases, we exclude the coins locked by the corrupt node while computing the capacity locked. We measure the difference in capacity locked in \emph{HTLC} and $\textrm{HTLC-GP}^{\zeta}$.

\setlength\abovedisplayshortskip{-1pt}
\begin{equation}
\begin{matrix}
    \frac{\mathcal{B}_{EX}}{L} (n-1)\alpha- \frac{\mathcal{B}_{EX}}{L} ((\tilde{n}^{\zeta,k}-1)v + \sum\limits_{i=1}^{\tilde{n}^{\zeta,k}-1}Z_{v,i})\\ 
\qquad =    \frac{\mathcal{B}_{EX}}{L} \Big( (n-1)\alpha- ((\tilde{n}^{\zeta,k}-1)v + \sum\limits_{i=1}^{\tilde{n}^{\zeta,k}-1}Z_{v,i}) \Big)\\
\qquad  =    \frac{\mathcal{B}_{EX}}{L} \Big( (n-1)v(1+\gamma^{\zeta,k}\sum\limits_{j=1}^{\tilde{n}^{\zeta,k}} t_j) -\\ ((\tilde{n}^{\zeta,k}-1)v + \sum\limits_{i=1}^{\tilde{n}^{\zeta,k}-1}Z_{v,i}) \Big) \\
    
\qquad  =    v\frac{\mathcal{B}_{EX}}{L} \Big( (n-\tilde{n}^{\zeta,k})+ \gamma^{\zeta,k} \tilde{n}^{\zeta,k} ((n-1)(D+\frac{(\tilde{n}^{\zeta,k}-1)\Delta}{2})\\ -\frac{\tilde{n}^{\zeta,k}-1}{2}(D+\frac{(2\tilde{n}^{\zeta,k}-1)\Delta}{3})) \Big) \\
    
    \end{matrix}
\end{equation}

The loss percent is ratio of difference of capacity locked in \emph{HTLC} and HTLC-GP$^{\zeta}$ and capacity locked in \emph{HTLC}.
\begin{equation}
\begin{matrix}
\frac{ v\frac{\mathcal{B}_{EX}}{L} \Big( (n-\tilde{n}^{\zeta,k})+ \gamma^{\zeta,k} \tilde{n}^{\zeta,k} ((n-1)(D+\frac{(\tilde{n}^{\zeta,k}-1)\Delta}{2})-\frac{\tilde{n}^{\zeta,k}-1}{2}(D+\frac{(2\tilde{n}^{\zeta,k}-1)\Delta}{3})) \Big)}{    \frac{\mathcal{B}_{EX}}{L} (n-1)\alpha} \\
\end{matrix}
\end{equation}

Considering $t_{n-1}=D$ and $t_i=D+(n-i-1)\Delta, i \in [0,n-1]$, where $\sum\limits_{j=0}^{n-1}t_j=nD+\frac{n(n-1)}{2}\Delta$ and substituting $\alpha=v(1+\gamma^{\zeta,k} \sum\limits_{j=0}^{n-1} t_j)$, the loss percent is
\begin{equation}
\begin{matrix}
\frac{ v\frac{\mathcal{B}_{EX}}{L} \Big( (n-\tilde{n}^{\zeta,k})+ \gamma^{\zeta,k} \tilde{n}^{\zeta,k} ((n-1)(D+\frac{(\tilde{n}^{\zeta,k}-1)\Delta}{2})-\frac{\tilde{n}^{\zeta,k}-1}{2}(D+\frac{(2\tilde{n}^{\zeta,k}-1)\Delta}{3})) \Big)}{    \frac{\mathcal{B}_{EX}}{L} (n-1)v(1+\gamma^{\zeta,k} \sum\limits_{j=0}^{n-1} t_j)} \\
\qquad= \frac{n-\tilde{n}^{\zeta,k}}{(n-1)\Big(1+\gamma^{\zeta,k} n D+ \gamma^{\zeta,k} n \Delta \frac{n-1}{2}\Big)}+\\ \frac{\gamma^{\zeta,k} \tilde{n}^{\zeta,k} ((n-1)(D+\frac{(\tilde{n}^{\zeta,k}-1)\Delta}{2})-\frac{\tilde{n}^{\zeta,k}-1}{2}(D+\frac{(2\tilde{n}^{\zeta,k})\Delta}{3}))}{(n-1)\Big(1+\gamma^{\zeta,k} n D+ \gamma^{\zeta,k} n \Delta \frac{n-1}{2}\Big)}

\end{matrix}
\end{equation}

\end{claimproof}

The loss percent is dominated by the term $\frac{n-\tilde{n}^{\zeta,k}}{n-1}$. For a given $k$, if $\zeta$ increases, the maximum path length $\textcolor{black}{\tilde{n}^{\zeta,k}}$ decreases, and so the loss incurred increases. In that case, the attacker would prefer to invest in other activities with higher returns rather than mount an attack on the network.

\section{Experimental analysis}
\label{experiment}
(a) \emph{Setup:} For our experiments, we use Python 3.8.2 and NetworkX, version 2.4 \cite{hagberg2008exploring}. System configuration used is Intel Core i5-8250U CPU, Operating System: Kubuntu 20.04, Memory: 7.7 GiB of RAM. We use twelve snapshots \cite{mizrahi2020congestion} of Bitcoin Lightning Network taken over a year, starting from \emph{September 2019}, have been used. Since our proposed strategy for countering griefing attacks requires both parties to fund the channel, we divide the channel's capacity into equal halves, allocating each half as the balance of a counterparty. 

%
%
%
%
%

\vspace*{0.1cm}
(b) \emph{Evaluation Methodology}: We define the strategy used by the attacker in the network. The latter corrupts the nodes that are either pendant vertices or have just one channel in the network. It is easier and more cost-effective to target such peripheral nodes than nodes with high centrality. A highly central node earns a higher profit as transactions tend to get routed through such nodes. Also, the attacker needs to offer a higher incentive per attack, which may not be a good strategy. On the other hand, peripheral nodes can be easily incentivized to deviate, as they haven't gained much trust in the network. Such nodes do not expect to earn much by behaving altruistically. While analyzing the effectiveness of \emph{HTLC-GP} and $\textrm{HTLC-GP}^{\zeta}$, we assume that $U_{i-1}$ always forward the payment request to $U_i$. 


The dataset and parameters used in our experiments are as follows: (i) \emph{HTLC-GP}: The set of experiments are divided into two parts. Transaction value is varied between $10000\textrm{-}100000$ satoshis and $\gamma \in $ [$10^{-7},10^{-3}$]. In the first part, we analyze the decrease in capacity locked when a penalty is introduced. The budget of the attacker is varied between $0.05 \ BTC-6.25 \ BTC$. The path length is set to $n=20$ and $D$ is set to 100. In the second part, we analyze the rate of the successful transaction in the absence of a griefing attack. We vary the number of transactions between 3000-9000 and path length $\kappa$ is varied between 5 and 20. (ii) $\textrm{HTLC-GP}^{\zeta}$: We analyze the further decrease in capacity locked upon introduction of penalty, as well as guaranteed minimum compensation in the payment protocol for a given budget of the attacker. The budget of the attacker is varied between 0.05 BTC-6.25 BTC. The transaction value is varied between $10000$ satoshis to $100000$ satoshis. We vary the parameter $k$ between $0.005$ to $2$. For a fixed $k$, $\zeta$ is varied so that path length ranges between $2$ to $20$. Both $D$ and $\Delta$ are set to 100.

\subsection{Observations} 
\begin{itemize}
\itemsep0em 

 \begin{figure}[t]

      \centering
     \subfloat[Capacity locked (in BTC) vs Adversary's Budget]{{\includegraphics[width=6cm]{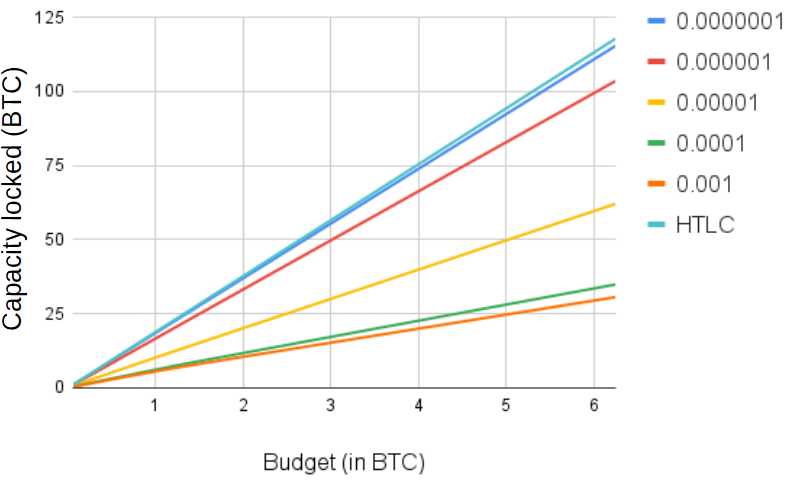} }}%
     \qquad
     \subfloat[Ratio of successful transaction (HTLC-GP/HTLC) upon varying $\gamma$]{{\includegraphics[width=6cm]{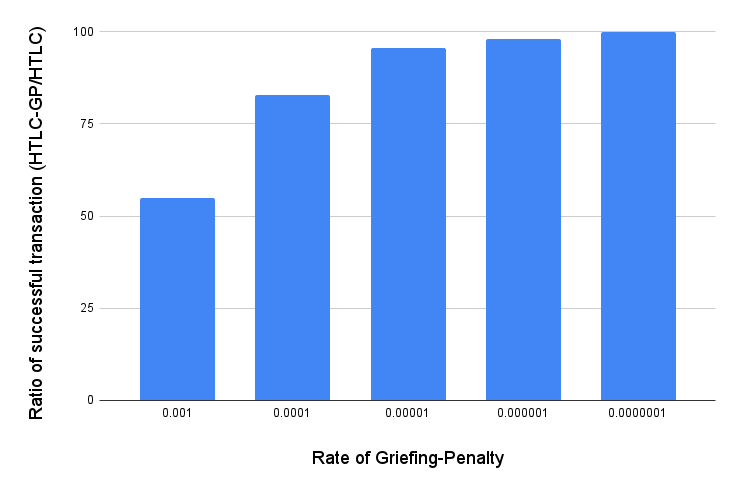} }}%
     \qquad
     \caption{$\gamma$ is varied between $10^{-3}$ to $10^{-7}$}
     \label{fig:penalty2}%
 \end{figure}

\item \emph{Effectiveness of HTLC-GP}: The plot in Fig.\ref{fig:penalty2}(a) shows that the capacity locked drops from 90\% to 20\% when $\gamma$ is varied between $10^{-7}$ to $10^{-3}$. We see a sharp decrease in capacity locked when $\gamma$ increases from $10^{-6}$ to $10^{-5}$, with the capacity locked dropping from $82\%$ to $50\%$. When $\gamma$ is $10^{-4}$, the capacity locked drops to $25\%$. The plot in Fig.\ref{fig:penalty2}(b) shows that the ratio of successful transaction executed drops to 54\% when $\gamma$ is $10^{-3}$ and it is around 99\% when $\gamma$ is $10^{-7}$.

\item \emph{Effectiveness of $\textrm{HTLC-GP}^{\zeta}$}: $k$ is varied between $0.005$ and $2$, and for each $k$, the factor $\zeta$ is varied so that the maximum path length ranges between 2 and 20. We observe that on varying $k$ and $\zeta$, $\gamma$ varies between $10^{-7}$ to $10^{-3}$. The drop in capacity locked in the network ranges between $18\%$ to $46\%$. The drop in capacity locked is significant for the lower value of $\gamma$ and the difference reduces for $\gamma>10^{-5}$. Table \ref{tab2} provides a comparative analysis of percentage loss in capacity between \emph{HTLC-GP} and $\textrm{HTLC-GP}^{\zeta}$.

%
%

\begin{table}
\begin{center}
\scalebox{0.5}{
 \begin{tabular}{|c| c |c|c|c|c|c|c|} 
 \hline
$k$ &$\zeta$ &$\gamma^{\zeta,k}$ &Maximum Path Length &$\gamma$ &Maximum Path Length & \multicolumn{2}{c|}{Ratio of capacity locked}\\[1ex]
\cline{7-8}
 & &$HTLC-GP^{\zeta}$ &$HTLC-GP^{\zeta} (n^{\zeta,k})$ &HTLC-GP & HTLC-GP ($n$) &$\frac{HTLC-GP^{\zeta}}{HTLC}$ &$\frac{HTLC-GP}{HTLC}$ \\ [1ex] 
 \hline
  0.005  &0.00025 &$2.4 \times 10^{-7}$ &20 &$2.4 \times 10^{-7}$ &20   &96.89\% &96.89\%\\
   &0.0005 &$9.1 \times 10^{-7}$ &10 &$9.1 \times 10^{-7}$ &20 &46.3\% &89.86\%\\
    &0.0025 &$1.4 \times 10^{-5}$ &2 &$1.4 \times 10^{-5}$ &20  &5.21\% &50.7\% \\
\hline

 0.01  &0.0005 &$4.7\times 10^{-7}$ &20 &$4.7\times 10^{-7}$ &20   &94\% &94\%\\
   &0.001 &$1.8\times 10^{-6}$ &10 &$1.8\times 10^{-6}$ &20  &44.8\% &81.5\%\\
    &0.005 &$2.8\times 10^{-5}$ &2 &$2.8\times 10^{-5}$ &20  &5.1\% &42\% \\
\hline
 0.05  &0.0025 &$2.4\times 10^{-6}$ &20 &$2.4\times 10^{-6}$ &20   &78\% &78\%\\
   &0.005 &$9.1\times 10^{-6}$ &10 &$9.1\times 10^{-6}$ &20  &38.2\% &54\%\\
    &0.025 &$1.6\times 10^{-4}$ &2 &$1.6\times 10^{-4}$ &20  &4.7\%\% &33\% \\
\hline
 0.1  &0.005 &$4.8\times 10^{-6}$ &20 &$4.8\times 10^{-6}$ &20   &67.5\% &67.5\%\\
   &0.01 &$1.8\times 10^{-5}$ &10 &$1.8\times 10^{-5}$ &20  &33.5\% &45.5\%\\
    &0.05 &$3.3\times 10^{-4}$ &2 &$3.3\times 10^{-4}$ &20  &4.45\% &32.5\% \\
\hline

 0.25  &0.0125 &$1.2\times 10^{-5}$ &20 &$1.2\times 10^{-5}$ &20   &53\% &53\%\\
   &0.025 &$4.5\times 10^{-5}$ &10 &$4.5\times 10^{-5}$ &20  &28\% &40\%\\
    &0.1125 &$6.9\times 10^{-4}$ &2 &$6.9\times 10^{-4}$ &20  &4.2\% &32\% \\
\hline
 0.5  &0.025 &$2.4\times 10^{-5}$  &20 &$2.4\times 10^{-5}$  &20  &44\% &44\%\\
  &0.05  &$9.1\times 10^{-5}$ &10 &$9.1\times 10^{-5}$ &20 &22.1\% &38.5\% \\
    &0.2 &$1.1\times 10^{-3}$ &2 &$1.1\times 10^{-3}$ &20 &3.8\% &31.5\%\\
\hline
 0.75  &0.0375 &$3.6\times 10^{-5}$ &20 &$3.6\times 10^{-5}$ &20  &41\%  &41\%\\
   &0.075     &$1.36\times 10^{-4}$ &10  &$1.36\times 10^{-4}$ &20 &21.2\% &35.6\% \\
    &0.3   &$1.7 \times 10^{-3}$ &2  &$1.7 \times 10^{-3}$ &20 &3.51\% &30.04\%\\
\hline

 1  &0.05 &$4.8\times 10^{-5}$ &20 &$4.8\times 10^{-5}$ &20   &38\% &38\%\\
   &0.1 &$1.8\times 10^{-4}$ &10 &$1.8\times 10^{-4}$ &20  &20\% &34\%\\
    &0.5 &$3.3\times 10^{-3}$ &2 &$3.3\times 10^{-3}$ &20   &3.4\% &29.98\%\\
\hline
 2  &0.1 &$10^{-4}$ &20 &$10^{-4}$ &20   &35\% &35\%\\
   &0.2 &$3.6\times 10^{-4}$ &10 &$3.6\times 10^{-4}$ &20  &18\% &32\%\\
    &0.95 &$6.1\times 10^{-3}$ &2 &$6.1\times 10^{-3}$ &20   &3.34\% &29.01\%\\
\hline

\end{tabular}
}
\end{center}

\caption{Capacity Locked when $k$ and $\zeta$ is varied}
\label{tab2}
\end{table}

\end{itemize}

\subsection{Discussions}
\begin{itemize}

%
\item \emph{HTLC-GP}:  If $\gamma$ increases, the net capacity locked by the attacker decreases but uncorrupt participants are forced to lock extra collateral for a given transaction. This results in a drop in the success rate of transactions being processed due to a lack of liquidity in channels.

\item $\textrm{HTLC-GP}^{\zeta}$: When $\gamma$ increases, percentage loss in capacity locked for \emph{HTLC-GP} increases as well. But this is at the cost of a high failure rate of transactions. In this protocol, we observe that the capacity locked drops substantially even for lower values of $\gamma$ when $k$ and $\zeta$ are adjusted to reduce the maximum path length. 

(i) $D$ is varied: The limit on the maximum timeout period in an \emph{HTLC} is 2016 blocks \cite{mizrahi2020congestion}. So $D$ cannot be increased indefinitely. $\gamma^{\zeta,k}$ will decrease if $D$ increases. The capacity locked remains invariant as the cumulative penalty does not change abruptly.  

(ii) $\zeta$ is varied: For a fixed value of $k$, $\zeta$ can be increased, reducing the maximum path length available for routing. This will increase the cost of the attack. When the majority of participants in the network adhere to non-attacking behavior, then the compensation offered can be reduced, readjusting the path length. Hence, the parameters must be chosen accordingly.

\end{itemize}

\section{Scalability Analysis}
\label{scale}
\textcolor{black}{We analyze the performance of \emph{HTLC, HTLC-GP} and $\textrm{HTLC-GP}^{\zeta}$ on the snapshots selected in Section \ref{experiment}. The number of transaction request is varied between 100 to 65000. Transaction chose the shortest feasible path as per the availability of capacity in the network so the path length varied between 4 to 12. For \emph{HTLC-GP} and $\textrm{HTLC-GP}^{\zeta}$, $\gamma$ is varied between $10^{-6}$ to $0.001$. We analyze the performance of the protocols in two different models - (i) All uncorrupt players are altruistic and (ii) All uncorrupt players are rational}

\subsubsection*{(i) Uncorrupt players are altruistic}
\begin{figure}[!ht]

      \centering
\includegraphics[width=7cm]{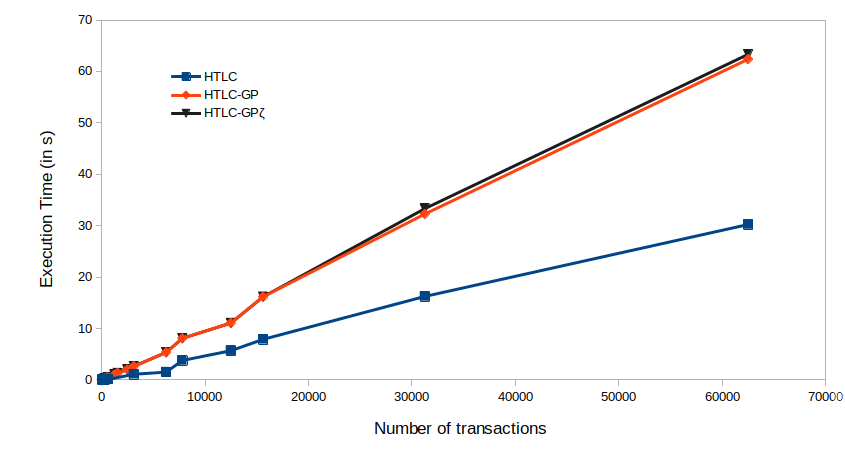}%
     \caption{Comparative analysis of execution time of \emph{HTLC, HTLC-GP}, and $\textrm{HTLC-GP}^{\zeta}$}
     \label{fig:scale}%
 \end{figure}

\textcolor{black}{All the nodes follows the step of protocol without assesing the risk. Taking an average over all the instances, we plot the dependency of execution time on the number of transaction request processed in the network in Fig. \ref{fig:scale}. We observe that the execution time of $\textrm{HTLC-GP}^{\zeta}$ is around 62s when number of transaction request is 65000. Since execution steps of \emph{HTLC-GP} and $\textrm{HTLC-GP}^{\zeta}$ are same, both takes the same time for execution for a given path length and rate of griefing-penalty. The execution time of \emph{HTLC} does not exceed 30s when the number of transaction request is 65000. The reason for an increase in execution time of \emph{HTLC-GP} (or $\textrm{HTLC-GP}^{\zeta}$) is that it becomes difficult after sometime to get a path with sufficient capacity for locking the payment as well as the penalty. Hence a node has to repeatedly search for neighbor that would have sufficient capacity to route payment.}

\subsubsection*{(ii) Uncorrupt players are rational} 

\textcolor{black}{ We varied the belief $\theta$ between 0 to 0.8. However, \emph{HTLC} fails to execute when $\theta>0.025$. None of the nodes in the network chooses to participate for the fear of loss due to griefing attack. This is not the case with \emph{HTLC-GP} (or $\textrm{HTLC-GP}^{\zeta}$). The protocol executes succesfully even for $\theta\leq 0.7$. The analysis of belief based on which a participant would be willing to forward payment is provied in Section III of Appendix. It is difficult to compare the run-time of the protocols when one of them fails to execute in a rational model. This shows that both \emph{HTLC-GP} and $\textrm{HTLC-GP}^{\zeta}$ are robust in a rational model. For a given channel, we computed the execution cost \cite{halpern2015algorithmic}\footnote{\textcolor{black}{As discussed in \cite{halpern2015algorithmic}, the cost incurred in a game model can be in terms of execution time or space or locking of extra coins in this context. We show that the execution time is negligible. However, an intermediate party has to lock additional coins in the form of penalty in \emph{HTLC-GP}. Opportunity coins of keeping additional coins unutilized have been taken into consideration while discussing the payoff structure of $\Gamma_{HTLC-GP}$.}} of setting up the game model and the time taken by a participant to decide whether to forward or not forward a payment. The execution time was of the order of microseconds. So the possibility of an additional cost incurred while setting up the game model between any two parties can be ruled out.}

\subsubsection*{Inference}
\textcolor{black}{The scalability analysis shows that the execution time of \emph{HTLC-GP} (or $\textrm{HTLC-GP}^{\zeta}$) is around twice of the run time of \emph{HTLC} in presence of altruistic players. However, players are rational and they will not follow the protocol blindly. We have assessed and shown in Fig. \ref{fig:penalty2}(a) that a griefing attack can prove to be fatal for the network in terms of unutilized coins. In a rational model, \emph{HTLC} fails even when the belief of a player being corrupt is as low as $0.025$. On the contrary, players are willing to participate in both \emph{HTLC-GP} or $\textrm{HTLC-GP}^{\zeta}$ even if the belief of a player being corrupt is as high as 0.7. Adding a penalization mechanism and controlling the maximum allowed path length for payment proves beneficial in curbing the impact of such attacks, safeguarding the interest of honest participants in the network.}

\section{Conclusion}
\label{conclusion}
In this paper, we perform a strategic analysis of griefing attacks in Lightning Network. We define a two-player game model where one party chooses its strategy based on its' belief of the other player's type. We have analyzed the effectiveness of payment protocol \emph{HTLC-GP} in the same model. We observe that the cost of attack increases with the introduction of the penalty. However, \emph{HTLC-GP} is found to be weakly effective in countering the attack as it is dependent on the rate of griefing penalty. To further increase the cost of the attack, we introduce the concept of guaranteed minimum compensation for the affected parties and control the maximum path length used for routing. We discuss a modified payment protocol $\textrm{HTLC-GP}^{\zeta}$, and our experimental results show that the former is more effective than \emph{HTLC-GP} in countering the griefing attack. As a part of our future work, we would like to analyze the impact of network congestion on the uncorrupt party's willingness to lock penalty and extend it to a multi-party game model. We would also like to analyze the griefing attack in the \emph{BAR model} \cite{aiyer2005bar} since inclusion of \emph{Byzantine} node will lead to a more realistic modeling of the attack in Bitcoin Network. Lastly, we would like to propose an incentive-compatible countermeasure in presence of rational miners. Tsabary et al. \cite{tsabary2021mad} had proposed \emph{MAD-HTLC} to counter bribery attack in \emph{HTLC} but they have not discussed griefing attack. Our objective would be to combine the best of the both world and propose a stronger protocol. 



\section*{Acknowledgement}
We thank the anonymous reviewers of IEEE TNSM for
their valuable feedback. We thank Dr. Stefanie Roos, Assistant Professor for distributed systems at TU Delft and the Delft Blockchain Lab, for her feedback on this work.  The work was partially supported by the European Research Council (ERC) under the European
Union’s Horizon 2020 research (grant agreement 771527-
BROWSEC), by the Austrian Science Fund (FWF) through
the projects PROFET (grant agreement P31621), and the project W1255-
N23, by the Austrian Research Promotion Agency (FFG)
through the COMET K1 SBA and COMET K1 ABC, by
the Vienna Business Agency through the project Vienna
Cybersecurity and Privacy Research Center (VISP), by the
Austrian Federal Ministry for Digital and Economic Affairs,
the National Foundation for Research, Technology and Development and the Christian Doppler Research Association
through the Christian Doppler Laboratory Blockchain Technologies for the Internet of Things (CDL-BOT).
\bibliographystyle{ieeetr}

\bibliography{PCN}

\begin{thebibliography}{10}

\bibitem{nakamoto2008bitcoin}
S.~Nakamoto, ``\textcolor{black}{Bitcoin: A peer-to-peer electronic cash
  system},'' {\em Decentralized Business Review}, p.~21260, 2008.

\bibitem{visa}
R.~Vlastelica, ``Why bitcoin won\'t displace visa or mastercard soon.''
  \url{https://www.marketwatch.com/story/why-bitcoin-wont-displace-visa-or-mastercard-soon-2017-12-15},
  December 2017.

\bibitem{croman2016scaling}
K.~Croman, C.~Decker, I.~Eyal, A.~E. Gencer, A.~Juels, A.~Kosba, A.~Miller,
  P.~Saxena, E.~Shi, E.~G. Sirer, {\em et~al.}, ``On scaling decentralized
  blockchains,'' in {\em International conference on financial cryptography and
  data security}, pp.~106--125, Springer, 2016.

\bibitem{gudgeon2019sok}
L.~Gudgeon, P.~Moreno-Sanchez, S.~Roos, P.~McCorry, and A.~Gervais, ``Sok:
  Layer-two blockchain protocols,'' in {\em International Conference on
  Financial Cryptography and Data Security}, pp.~201--226, Springer, 2020.

\bibitem{decker2015fast}
C.~Decker and R.~Wattenhofer, ``A fast and scalable payment network with
  bitcoin duplex micropayment channels,'' in {\em Symposium on Self-Stabilizing
  Systems}, pp.~3--18, Springer, 2015.

\bibitem{poon2016bitcoin}
J.~Poon and T.~Dryja, ``The bitcoin lightning network: Scalable off-chain
  instant payments,'' 2016.

\bibitem{egger2019atomic}
C.~Egger, P.~Moreno-Sanchez, and M.~Maffei, ``Atomic multi-channel updates with
  constant collateral in bitcoin-compatible payment-channel networks,'' in {\em
  Proceedings of the 2019 ACM SIGSAC Conference on Computer and Communications
  Security}, pp.~801--815, 2019.

\bibitem{robinson2019htlcs}
D.~Robinson, ``Htlcs considered harmful,'' in {\em Stanford Blockchain
  Conference}, 2019.

\bibitem{bank}
Z.~Lu, R.~Han, and J.~Yu, ``\textcolor{black}{Bank run Payment Channel
  Networks},'' {\em IACR Cryptol. ePrint Arch.}, vol.~2020, p.~456, 2020.

\bibitem{lu2020general}
Z.~Lu, R.~Han, and J.~Yu, ``General congestion attack on htlc-based payment
  channel networks,'' in {\em 3rd International Conference on Blockchain
  Economics, Security and Protocols (Tokenomics 2021)}, 2021.

\bibitem{rohrer2019discharged}
E.~Rohrer, J.~Malliaris, and F.~Tschorsch, ``Discharged payment channels:
  Quantifying the lightning network's resilience to topology-based attacks,''
  in {\em 2019 IEEE European Symposium on Security and Privacy Workshops
  (EuroS\&PW)}, pp.~347--356, IEEE, 2019.

\bibitem{agrief}
S.~{Mazumdar}, P.~{Banerjee}, and S.~{Ruj}, ``Time is money: Countering
  griefing attack in lightning network,'' in {\em 2020 IEEE 19th International
  Conference on Trust, Security and Privacy in Computing and Communications
  (TrustCom)}, pp.~1036--1043, 2020.

\bibitem{malavolta2019anonymous}
G.~Malavolta, P.~Moreno-Sanchez, C.~Schneidewind, A.~Kate, and M.~Maffei,
  ``\textcolor{black}{Anonymous Multi-Hop Locks for Blockchain Scalability and
  Interoperability},'' in {\em 26th Annual Network and Distributed System
  Security Symposium, NDSS 2019}, 2019.

\bibitem{rogaway2004cryptographic}
P.~Rogaway and T.~Shrimpton, ``Cryptographic hash-function basics: Definitions,
  implications, and separations for preimage resistance, second-preimage
  resistance, and collision resistance,'' in {\em International workshop on
  fast software encryption}, pp.~371--388, Springer, 2004.

\bibitem{gibbons1992dynamic}
R.~S. Gibbons, ``Dynamic games of complete information,'' in {\em Game Theory
  for Applied Economists}, pp.~55--142, Princeton University Press, 1992.

\bibitem{fudenberg1991perfect}
D.~Fudenberg and J.~Tirole, ``Perfect bayesian equilibrium and sequential
  equilibrium,'' {\em journal of Economic Theory}, vol.~53, no.~2,
  pp.~236--260, 1991.

\bibitem{zappala2020game}
P.~Zappal{\`a}, M.~Belotti, M.~Potop-Butucaru, and S.~Secci, ``Game theoretical
  framework for analyzing blockchains robustness,'' in {\em Proceedings of the
  4th International Symposium on Distributed Computing, Leibniz International
  Proceedings in Informatics (LIPIcs), Freiburg (virtual conference), Germany},
  pp.~49:1--49:3, 2020.

\bibitem{rain2021towards}
S.~Rain, Z.~Avarikioti, L.~Kov{\'a}cs, and M.~Maffei,
  ``\textcolor{black}{Towards a Game-Theoretic Security Analysis of Off-Chain
  Protocols},'' in {\em 36th IEEE Computer Security Foundations Symposium (CSF)
  (pp. nn-nn). IEEE Computer Society, Washington, DC, USA}, 2023.

\bibitem{xu2020game}
J.~Xu, D.~Ackerer, and A.~Dubovitskaya, ``A game-theoretic analysis of
  cross-chain atomic swaps with htlcs,'' in {\em 2021 IEEE 41st International
  Conference on Distributed Computing Systems (ICDCS)}, pp.~584--594, IEEE,
  2021.

\bibitem{han2019optionality}
R.~Han, H.~Lin, and J.~Yu, ``On the optionality and fairness of atomic swaps,''
  in {\em Proceedings of the 1st ACM Conference on Advances in Financial
  Technologies}, pp.~62--75, 2019.

\bibitem{russell}
``A proposal for up-front payments.''
  \url{https://lists.linuxfoundation.org/pipermail/lightning-dev/2019-November/002282.html},
  November 2019.

\bibitem{zmn}
``Proof-of-closure as griefing attack mitigation.''
  \url{https://lists.linuxfoundation.org/pipermail/lightning-dev/2020-April/002608.html},
  April 2020.

\bibitem{danezis2009sphinx}
G.~Danezis and I.~Goldberg, ``Sphinx: A compact and provably secure mix
  format,'' in {\em 2009 30th IEEE Symposium on Security and Privacy},
  pp.~269--282, IEEE, 2009.

\bibitem{azouvi2019sok}
S.~Azouvi and A.~Hicks, ``\textcolor{black}{SoK: Tools for {Game} {Theoretic}
  {Models} of {Security} for {Cryptocurrencies}},'' {\em Cryptoeconomic
  Systems}, vol.~0, apr 5 2021.
\newblock https://cryptoeconomicsystems.pubpub.org/pub/azouvi-sok-security.

\bibitem{garay2013rational}
J.~Garay, J.~Katz, U.~Maurer, B.~Tackmann, and V.~Zikas, ``Rational protocol
  design: Cryptography against incentive-driven adversaries,'' in {\em 2013
  IEEE 54th Annual Symposium on Foundations of Computer Science}, pp.~648--657,
  IEEE, 2013.

\bibitem{aiyer2005bar}
A.~S. Aiyer, L.~Alvisi, A.~Clement, M.~Dahlin, J.-P. Martin, and C.~Porth,
  ``Bar fault tolerance for cooperative services,'' in {\em Proceedings of the
  twentieth ACM symposium on Operating systems principles}, pp.~45--58, 2005.

\bibitem{buchanan1991opportunity}
J.~M. Buchanan, ``Opportunity cost,'' in {\em The world of economics},
  pp.~520--525, Springer, 1991.

\bibitem{osuntokun2020lightning}
O.~Osuntokun, C.~Fromknecht, W.~Paulino, O.~Gugger, and J.~Halseth, ``Lightning
  pool: A non-custodial channel lease marketplace,'' 2020.

\bibitem{gebraselase2021transaction}
B.~G. Gebraselase, B.~E. Helvik, and Y.~Jiang, ``Transaction characteristics of
  bitcoin,'' in {\em 2021 IFIP/IEEE International Symposium on Integrated
  Network Management (IM)}, pp.~544--550, IEEE, 2021.

\bibitem{gebraselase2021analysis}
B.~G. Gebraselase, B.~E. Helvik, and Y.~Jiang, ``An analysis of transaction
  handling in bitcoin,'' in {\em 2021 IEEE International Conference on Smart
  Data Services (SMDS)}, pp.~162--172, IEEE, 2021.

\bibitem{guasoni2021lightning}
P.~Guasoni, G.~Huberman, and C.~Shikhelman, ``Lightning network economics:
  Channels,'' {\em Available at SSRN 3840374}, 2021.

\bibitem{kawase2017transaction}
Y.~Kawase and S.~Kasahara, ``Transaction-confirmation time for bitcoin: A
  queueing analytical approach to blockchain mechanism,'' in {\em International
  Conference on Queueing Theory and Network Applications}, pp.~75--88,
  Springer, 2017.

\bibitem{fee1}
``Lightning 101: Lightning network fees.''
  \url{https://blog.bitmex.com/the-lightning-network-part-2-routing-fee-economics/},
  2019.
\newblock Accessed: 2019-01-22.

\bibitem{fee2}
``The lightning network (part 2) - routing fee economics.''
  \url{https://blog.bitmex.com/the-lightning-network-part-2-routing-fee-economics/},
  2019.
\newblock Accessed: 2019-03-27.

\bibitem{narahari2012game}
Y.~Narahari, {\em Game theory and mechanism design}, vol.~4.
\newblock World Scientific, 2014.

\bibitem{timelock}
BtcDrak, M.~Friedenbach, and E.~Lombrozo, ``Bip 112, checksequenceverify.''
  \url{https://github.com/bitcoin/bips/blob/master/bip-0112.mediawiki},
  2015-08-10.

\bibitem{mizrahi2020congestion}
A.~Mizrahi and A.~Zohar, ``\textcolor{black}{Congestion attacks in payment
  channel networks},'' in {\em International Conference on Financial
  Cryptography and Data Security}, pp.~170--188, Springer, 2021.

\bibitem{beres2019cryptoeconomic}
F.~B{\' e}res, I.~A. Seres, and A.~A. Bencz{\' u}r, ``(v1) {A} {Cryptoeconomic}
  {Traffic} {Analysis} of {Bitcoin}\textquoteright{}s {Lightning} {Network},''
  {\em Cryptoeconomic Systems}, jun 22 2020.
\newblock https://cryptoeconomicsystems.pubpub.org/pub/b8rb0ywn.

\bibitem{goldschlag1999onion}
D.~Goldschlag, M.~Reed, and P.~Syverson, ``Onion routing,'' {\em Communications
  of the ACM}, vol.~42, no.~2, pp.~39--41, 1999.

\bibitem{hagberg2008exploring}
A.~Hagberg, P.~Swart, and D.~S~Chult, ``Exploring network structure, dynamics,
  and function using networkx,'' tech. rep., Los Alamos National Lab.(LANL),
  Los Alamos, NM (United States), 2008.

\bibitem{halpern2015algorithmic}
J.~Y. Halpern and R.~Pass, ``Algorithmic rationality: Game theory with costly
  computation,'' {\em Journal of Economic Theory}, vol.~156, pp.~246--268,
  2015.

\bibitem{tsabary2021mad}
I.~Tsabary, M.~Yechieli, A.~Manuskin, and I.~Eyal, ``Mad-htlc: because htlc is
  crazy-cheap to attack,'' in {\em 2021 IEEE Symposium on Security and Privacy
  (SP)}, pp.~1230--1248, IEEE, 2021.

\end{thebibliography}

\end{document}